\documentclass[aps,prx,twocolumn,superscriptaddress,natlib,showpacs]{revtex4-1}
\usepackage{natbib}
\usepackage{graphicx}
\usepackage{dcolumn}
\usepackage{bm}
\usepackage{color}
\usepackage{amssymb}

\begin{document}

\title{Persistent Sinai type diffusion in Gaussian random potentials with decaying 
spatial correlations}
\author{Igor Goychuk}
\email{igoychuk@uni-potsdam.de, corresponding author}
\affiliation{Institute of Physics and Astronomy, University of Potsdam, 
Karl-Liebknecht-Str. 24/25, 14476 Potsdam-Golm, Germany}
\author{Vasyl O. Kharchenko}
\affiliation{Institute of Applied Physics, Natl. Acad. Sci. Ukraine, 
58 Petropavlovskaya str., 40030 Sumy, Ukraine}
\author{Ralf Metzler}
\affiliation{Institute of Physics and Astronomy, University of Potsdam, 
Karl-Liebknecht-Str. 24/25, 14476 Potsdam-Golm, Germany}

\date{\today}

\begin{abstract}
Logarithmic or Sinai type subdiffusion is usually associated with random force
disorder and non-stationary potential fluctuations whose root mean squared
amplitude grows with distance. We show here that extremely persistent,
macroscopic ultraslow logarithmic diffusion also universally emerges at
sufficiently low temperatures in stationary Gaussian random potentials
with spatially decaying correlations, known to exist in a broad range of
physical systems. Combining results from extensive simulations with a scaling
approach we elucidate the physical mechanism of this unusual subdiffusion. In
particular, we explain why with growing temperature and/or time a first
crossover occurs to standard, power-law subdiffusion, with a time-dependent
power law exponent, and then a second crossover occurs to normal diffusion with
a disorder-renormalized diffusion coefficient. Interestingly, the initial,
nominally ultraslow diffusion turns out to be much faster than the universal
de Gennes-B\"assler-Zwanzig limit of the renormalized normal diffusion, which
physically cannot be attained at sufficiently low temperatures and/or for
strong disorder. The ultraslow diffusion is also non-ergodic and displays
a local bias phenomenon. Our simple scaling theory not only explains our
numerical findings, but qualitatively has also a predictive character.
\end{abstract}


\maketitle

\section{Introduction}

Systems with Gaussian energy disorder characterized by spatially decaying
correlations are ubiquitous in physics thanks to the central limit theorem
\cite{Bouchaud1990,BouchaudAnnPhys90}. Such systems include, for instance, disordered organic
photoconductors with long-range electrostatic interactions \cite{BasslerReview,
DunlapPRL96}, supercooled liquids \cite{BasslerPRL87}, as well as naturally
occurring DNA macromolecules encoding biological information in living systems
\cite{GerlandPNAS,LassigReview,SlutskyPRE}. Colloidal systems in quenched, random
laser-created potentials have also recently become experimentally available
\cite{EversReview,HanesJPCM,HanesPRE,BewerungePRA}.

A common line of thinking  \cite{HanggiRevModPhys} treats diffusion and transport
phenomena in such systems as normal on experimentally relevant scales, with a
potential disorder-renormalized diffusion coefficient $D_{\rm ren}=D_0\exp(-[
\beta\sigma]^2)$, where $D_0$ is the free diffusion coefficient of the diffusing
particle in absence of the potential, $\sigma=\langle \delta U^2(x)\rangle^{1/2}$
is the root-mean-squared amplitude of the potential fluctuations $\delta U(x)$,
and $\beta=1/(k_BT)$ is the Boltzmann factor proportional to inverse temperature.
This famed renormalization result by De Gennes \cite{DeGennes75}, B{\"a}ssler
\cite{BasslerPRL87}, and Zwanzig \cite{ZwanzigPNAS} corresponds to a common
measurable temperature dependence of transport coefficients in disordered glassy
systems which is almost indistinguishable from the Vogel-Fulcher-Tammann law
\cite{HecksherNatPhys}, another commonly used temperature dependence used to fit
experimental data.

It came as a surprise when extensive simulations of stochastic Langevin dynamics
in Gaussian random potentials with decaying spatial correlations by Romero and Sancho
\cite{Romero98} demonstrated that the emerging diffusion is anomalously slow over the 
whole time range of simulations and
characterized by a power law scaling $\langle\delta x^2(t)\rangle\propto t^\alpha$ ($0<\alpha
<1$) of the mean squared displacement.
This result is very remarkable indeed because within the mean field approximation it
is easy to show that the Gaussian energy disorder yields residence time distributions
$\psi(\tau)$ of a log-normal type in finite spatial trapping domains, all moments
of which are finite. This behavior is fundamentally different from the situation
of annealed exponential energetic disorder yielding residence time distributions
of the form $\psi(\tau)\propto1/\tau^{1+\alpha}$ with $\alpha=k_BT/\sigma$
\cite{Hughes}: Here subdiffusion emerges for $k_BT<\sigma$ because of the divergent
mean residence times, which provides the standard model of subdiffusion in the
continuous time random walk framework \cite{Hughes,Metzler01}.

Hence, Gaussian energy disorder cannot yield subdiffusion in the absence of spatial
correlations. To neglect such correlations is a usual procedure when they are short
ranged \cite{HanggiRevModPhys}. Whether organic photoconductors and other disordered
materials are better described by Gaussian or exponential models of the energy
disorder is a subject of continuing controversy. On one hand, the Gaussian model is
generally accepted for organic photoconductors \cite{BasslerReview}. On the other
hand, some recent experiments \cite{SchubertPRB13} seem to be more consistent with
the model of exponential disorder, leading to a similar time dependence of transient
currents as in amorphous semiconductor films \cite{Scher75}. Models of Gaussian
disorder with significant spatial correlations can be a key to resolve this
controversy. Physically, for instance, such short range correlations may correspond
to the size of a protein molecule diffusing over the potential landscape of a DNA
strand \cite{GoychukPRL14}, or to the size of a colloidal particle in an
uncorrelated laser field potential \cite{HanesJPCM}. 
Morever, long-range electrostatic interactions
give rise to long-range spatial correlations in organic photoconductors \cite{DunlapPRL96},
and biological information encoded in the sequence of DNA base pairs manifests itself in 
long-range base pair correlations \cite{PengNature,AvrahamHavlin}.

The remarkable discovery that subdiffusion persists beyond the mean field
approximation was reaffirmed in a number of recent studies \cite{KhouryPRL,
SimonPRE13,HanesPRE,GoychukPRL14}. It was explained recently in terms of a
finite non-ergodicity length $L_{\rm erg}$ of the spatial random process
$\exp[\beta\delta U(x)]$ \cite{GoychukPRL14} related to the presence of
spatial potential correlations. In this respect, the origin of a weak
ergodicity breaking---the fundamental difference between ensemble and time
averaged physical observables even in the limit of long observations times
\cite{Bouchaud92,Bel05,MetzlerPCCP}---is very different from the one in
systems described by continuous time random walks with annealed, uncorrelated
residence time distributions with divergent mean residence times \cite{Bouchaud92,
Bel05,BurovPRL07,LubelskiPRL08,HePRL08,Sokolov09}. For systems with Gaussian
disorder, the non-ergodic behavior occurs on the transient spatial scale $L_{\rm
erg}$ which, however, rapidly grows with $\beta\sigma$ for spatially correlated
potential fluctuations $\delta U(x)$. Indeed, this spatial scale can readily reach
macroscopic sizes already for disorder strengths $\sigma\sim4\div5k_BT$, which is
a typical value for disordered organic photoconductors already at room temperatures
\cite{DunlapPRL96}.

Subdiffusion occurs on the corresponding non-ergodic and generally mesoscopic
spatial scales. Somewhat surprisingly, diffusion retains its asymptotically normal
form for any decaying correlations, no matter how slow they decay
\cite{GoychukPRL14}.  However, for mesoscopic and even macroscopic systems this
limit can be out of reach and become physically completely irrelevant, leading
us to consider this type of subdiffusion as one of the fundamental processes in
a large range of applications. Some of these remarkable features have recently
been confirmed experimentally in colloidal systems \cite{HanesPRE} and, more
tentatively, also for diffusion of regulatory proteins on DNA strands
\cite{WangPRL}. Namely, although in Ref.~\cite{WangPRL} the results are
interpreted in terms of normal diffusion, the single-trajectory diffusion
coefficient is scattered over three orders of magnitude, which likely implies
a lack of ergodicity. The real physical mechanism of anomalous diffusion in
spatially correlated Gaussian energetic disorder remains elusive, and it is
the main goal of this paper to clarify its physical origin. The whole setup is
very different from the original scenario of Sinai diffusion \cite{Sinai82,Bouchaud1990,BouchaudAnnPhys90,Doussal99}.
It is also different from Sinai diffusion emerging from a
power law energy disorder \cite{AvrahamHavlin}, or logarithmic tails of the waiting time distribution \cite{Godec14} in the mean field approximation.  Notwithstanding,
we will show below that a generalized Sinai subdiffusion indeed emerges at
sufficiently low temperatures with $\langle\delta x^2(t)\rangle\propto[\ln(t/
t_0)]^a$, with a general scaling exponent $a$, which is $a=4$ in the special
case of Sinai diffusion \cite{Sinai82,Bouchaud1990,BouchaudAnnPhys90,Doussal99}. 
The latter one is usually
associated with the model of random force disorder \cite{Bouchaud1990,BouchaudAnnPhys90,
AvrahamHavlin,Doussal99} yielding potential fluctuations whose root mean square amplitude
exhibits an unlimited growth with distance, that is, the potential presents a
free Brownian motion in space \cite{Bouchaud1990,BouchaudAnnPhys90,Doussal99}. A 
generalized Sinai diffusion with $a\neq 4$ emerges also in the case of
fractional Brownian motion in space \cite{OshaninPRL}.

In this work, we reveal that a generalized Sinai type diffusion  universally
emerges for a sufficiently low temperature and/or strong disorder when $\beta
\sigma$ exceeds a typical value of 5 to 10, even for the case of very short range
exponential or linearly decaying correlations. This fundamental result is fully
confirmed by extensive simulations. Based on the arguments developed herein it
it should be absolutely possible to observe macroscopic Sinai diffusion at
experimentally relevant time scales, in systems governed by spatially correlated
Gaussian energetic disorder.

The remainder of this manuscript is structured as follows. In Section II we
develop the general concepts of our approach and introduce a scaling theory for
the resulting diffusive dynamics. Section III presents results from extensive
simulations for various relevant cases of the spatial correlations decay of the
potential fluctuations. In Section IV we present a detailed discussion of our
results and draw our conclusions.

\section{Basic concepts and scaling theory}

\subsection{Growing potential fluctuations}

We start from an observation, which is central to the theory of extreme
events \cite{Castillo} and pivotal for the present study. Namely, even if
a stationary Gaussian process $U(x)$ has a finite root mean squared fluctuation 
$\sigma$,
the maximal amplitude of its fluctuations $\delta U_{\rm max} (x)$ grows slowly
with the distance $x$, at odds with intuition \cite{Pickands69,ZhangPRL}. The
law of this growth can be found from the following consideration. We consider
the potential as a random process with zero average, $\langle U(x)\rangle=0$,
and ask the question of how often an energy level $U_{\rm max}$ is crossed with
growing distance $x\to\infty$, in a stationary limit. For any stationary,
differentiable Gaussian process with normalized autocorrelation function $g(x)=
\langle U(x+x')U(x')\rangle/\sigma^2$ and asymptotically decaying correlations,
the answer is known: The averaged number of level-crossing events $\langle n(x)
\rangle$ grows linearly with distance: $\langle n(x)\rangle=\Gamma x$, at the
rate $\Gamma=\sqrt{-g''(0)}\exp(-U_{\rm max}/2\sigma^2)/\pi$ \cite{Papoulis}.
Thus, the level $U_{\rm max}$ is typically crossed at the distance $x$ found
from $\langle n(x)\rangle=1$. Hence,
\begin{eqnarray}\label{first}
U_{\rm max}(x)=\sigma [2\ln (x/x_0) ]^{1/2},
\end{eqnarray}
where $x_0=\pi/\sqrt{-g''(0)}$. This is a very general and universal result
valid for any correlations with finite $g''(0)$, in agreement with the
results of Ref.~\cite{Pickands69}. Note that this result holds only
asymptotically, given a stationary regime is established. Moreover, the longer
the correlations range, the later this asymptotic result is established.

For example, for a Gaussian decay of the correlations we will have $g(x)=\exp
(-x^2/\lambda^2)$ with a correlation length scale parameter 
$\lambda$, and in this case $x_0=\pi\lambda/\sqrt{2}$.
Note that in a scaling sense such a correlation length parameter $\lambda$
always exists, even if the correlation length formally defined as $\lambda_{\rm
corr}=\int_0^{\infty}|g(x)|dx$ may diverge. The length $\lambda$ plays a
fundamental role in the presence of correlations. It can be, for instance,
of the size of the protein-DNA contact, or the size of the colloidal particle
in an uncorrelated laser field.

For power law decaying correlations with $g(x)=1/[1+x^2/\lambda^2]^{\gamma/2}$,
and $g(x)\propto 1/x^\gamma$ for $x\gg\lambda$, and thus $x_0=\pi\lambda/\sqrt
\gamma$. This case is very interesting in applications to diffusion of regulatory
proteins on DNA strands, 
where the real base pair sequence arrangement can play a very profound role \cite{BenichouPRL09,Bauer15}.
In particular, longe-range correlations in the base pair sequence,
which encode biological information, can indeed decay algebraically slow, as shown,
for instance, in Refs.~\cite{PengNature,AvrahamHavlin} with $\gamma\sim 0.6\div0.8$.
This should yield corresponding correlations in the protein binding energy with
$\lambda$ being of the size of the protein-DNA contact, or larger. Namely, we here
have a case for which $\lambda_{\rm corr}=\infty$, which suggests that the
subdiffusive regime should hold much longer than in the case of short ranged
correlations. This is indeed confirmed in our detailed analysis below. Another
important example is provided by some organic photoconductors, for which $\gamma=1$
and $\sigma=0.1\div0.125\;eV$ or $4\div5\;k_BT$ at room temperature
\cite{DunlapPRL96}.

The next case of relevance we consider here is the Ornstein-Uhlenbeck
process with $g(x)=\exp(-|x|/\lambda)$. This case may at first appear problematic
as it is not differentiable, and the corresponding force $f(x)=-\partial V(x)/
\partial x$ has the infinite root mean squared amplitude $\langle f^2(x)\rangle^{
1/2}=\infty $. However, in physical applications we consider in fact a regularized
Ornstein-Uhlenbeck process on a lattice with a finite grid size $\Delta x$. This
allows also for a numerical treatment of the corresponding continuous dynamics
(which otherwise would not be possible due to the infinite force root mean squared
amplitude). We note that this point is similar to the subtleties arising when we
consider the model of white Gaussian processes $f(x)$ in the theory of continuous
space Sinai diffusion \cite{Bouchaud1990}. Indeed, we always generate realizations
of Gaussian processes with some finite resolution $\Delta x$, following a well
established algorithm \cite{SimonFNL}. The corresponding $g(x)$ is getting, in
fact, thereby regularized. An appropriate formal regularization reads $g(x)=\exp(
-\sqrt{x^2+(\Delta x/2)^2}/\lambda+\Delta x/(2\lambda))$. It yields the same force
root mean squared $\langle f^2(0)\rangle^{1/2}=\sqrt{2}\sigma/\sqrt{\Delta x
\lambda}$ as the Ornstein-Uhlenbeck process on a lattice with grid size $\Delta x$.
In this case, $x_0=\pi(\lambda\Delta x/2)^{1/2}$. An important further remark is
that, when $\Delta x\to0$, then $x_0\to0$, and $U_{\rm max}(x)$ grows accordingly
for any fixed $x$. However, this growth is logarithmically slow, and for realistic
variations of $\Delta x$ the effect is small. Nevertheless, the case of
exponentially decaying correlations is especially interesting in the present
context. Notice that such short range correlations appear also in the case
of a biologically meaningless and completely random DNA because of a finite
size $\lambda$ of the protein-DNA contact, which is typically from 5 to 30
bp \cite{StewartGenetics}.

Another pertinent correlation model for the particular application of diffusion
on DNA is the model of linearly decaying correlations $g(x)=(1-|x|/\lambda)$ for
$|x|\leq \lambda$, and zero otherwise \cite{Yaglom}. It emerges when the protein
interacts locally with only one corresponding bp, in the immediate contact. The
corresponding random process $U(x)$ is also non-differentiable, and with respect
to the $U_{\rm max}(x)$ behavior its regularized version behaves rather similar
to that of the Ornstein-Uhlenbeck process (considered with the same $\Delta x$).

\begin{figure}
\includegraphics[width=8cm]{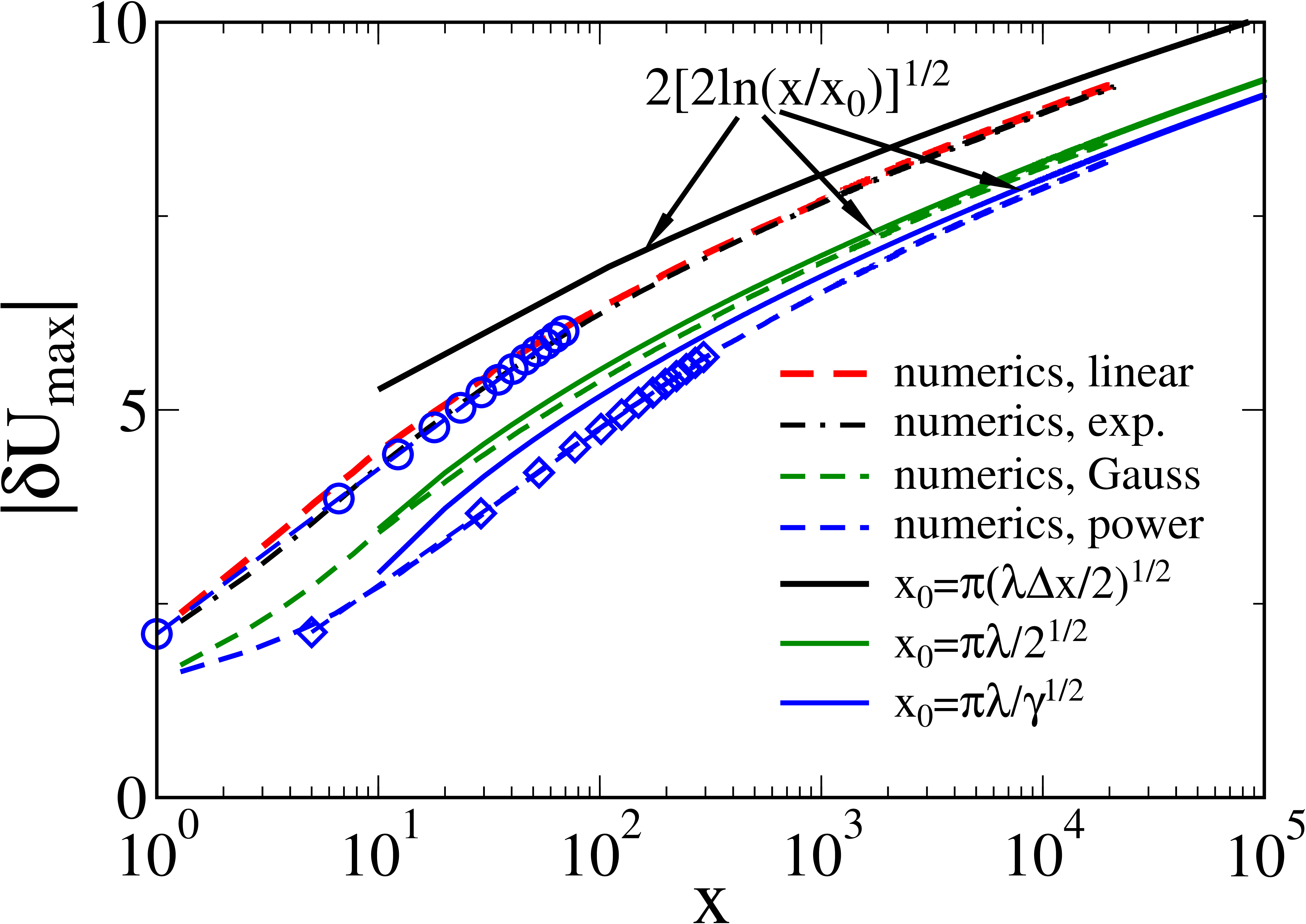}
\caption{Maximal amplitude of potential fluctuations $|\delta U_{\rm max}(x)|$
(in units of root mean squared fluctuations $\sigma$) starting from some random
point regarded as the origin, plotted versus distance $x$ (in units of the
correlation length $\lambda$) for several models of Gaussian potentials with
decaying correlations. This decay is (i) exponential, (ii) linear, (iii) Gaussian,
and (iv) of power law form with $\gamma=0.8$. Broken lines represent the results
from simulations and full lines depict the corresponding theoretical results
$2U_{\rm max}(x)$ from Eq.~(\ref{first}), which are gradually approached for long
distances $x$. In all cases, $\Delta x=0.02$ is chosen. The symbols show a $\log x$
scaling fit for either initial or intermediate values of $x$ in the case of
exponential and power law models of the correlations, respectively.}
\label{Fig1}
\end{figure} 

Let us check the above exact analytical results against the results from
extensive simulations: Fig.~\ref{Fig1} shows how the amplitude of the potential
fluctuations grows with distance for several models of Gaussian disorder. The
universal result in Eq.~(\ref{first}) agrees well with the numerics for all four
models considered here. A similar scaling was also found numerically for a
different model of correlations \cite{HanesJPCM}. The agreement is better for
differentiable (in the limit $\Delta x\to 0$) processes, and the model of linearly
decaying correlations behaves indeed rather similar to the Ornstein-Uhlenbeck model.
Notice that in the case of power law decaying correlations with $\gamma=0.8$, the
convergency is slower than for the Gaussian decay, and for much smaller $\gamma$,
for instance, $\gamma=0.2$, it is is still far from being achieved for the largest
$x$ in Fig.~\ref{Fig1} (not shown). The results in Fig.~\ref{Fig1} are obtained
from a well-known spectral method \cite{SimonFNL} given the corresponding
correlation functions, on a lattice with grid size $\Delta x$ ($\Delta x=0.02$ in
Fig.~\ref{Fig1}) and a maximal spatial interval $L_{\rm max}$ ($L_{\rm max}=N\Delta
x, N=2^{20}$ in Fig.~\ref{Fig1}). Notice also that the exact behavior of the
function $g(x)$ is used to generate the realizations of the Ornstein-Uhlenbeck
process and the process with linearly decaying correlations. The shown results
are obtained from averaging over $10^3$ uniformly distributed points (particles)
in each potential realization, and $10^2$ such potential realizations for
each model of correlations were taken.

It is clear that the strongest potential fluctuations occur in the case of
exponential and linear correlations. They are further increasing for smaller
values of the resolution $\Delta x$ due to the singularity of these models.
Another salient feature is that another scaling can be observed transiently.
Namely, we initially find the behavior $\log x$ rather than the $\sqrt{\log
x}$ scaling for the exponential and linear correlations, and intermediately
for the case of power law decaying correlations. Moreover, we notice an initial
$(\log x)^\delta$ scaling with $\delta>1$ for the power law and Gaussian decay.
To account for these various behaviors we henceforth consider the following
model
\begin{eqnarray}
\label{second}
|\delta U_{\rm max}(x)|=\sigma_{\rm eff}[\ln (x/x_{\rm in}) ]^{\delta}
\end{eqnarray}
for the potential fluctuations, where $\sigma_{\rm eff}$, $\delta$, and $x_{\rm
in}$ are fitting parameters, which are generally different from the theoretical
values $\sigma_{\rm eff}=2\sqrt{2}\sigma\approx 2.83\;\sigma $, $\delta=1/2$,
and $x_{\rm in}=x_0$ in Fig.~\ref{Fig1}. We will now use this generic empirical
formula and pursue the scaling argumentation used originally in the case of
continuous space Sinai diffusion \cite{Bouchaud1990}.

\subsection{Scaling theory}

Let us estimate the time $t$ a particle needs to travel the distance $x$ starting
at $x_0$, limited in an Arrhenius manner by the largest barrier met on its way,
$t=t_0\exp[\beta|\delta U_{\rm max}(x)|]$, where $t_0$ is a proportionality factor of
physical unit of time. From this scaling ansatz, in combination with relation
(\ref{second}) we immediately obtain our central result for the mean squared
displacement of the diffusing particle,
\begin{eqnarray}
\label{central}
\langle \delta x^2(t) \rangle = x_{\rm in} ^2 
\left\{e^{[(k_BT/\sigma_{\rm eff})\ln(t/t_0)]^{1/\delta}}-1 \right\}^2.
\end{eqnarray}
This in turn yields the asymptotic form
\begin{eqnarray}
\label{genSinai}
\langle \delta x^2(t) \rangle \approx 
 x_{\rm in}^2 
\left [(k_B T/\sigma_{\rm eff})\ln(t/t_0) \right ]^{2/\delta}
\end{eqnarray}
for small temperatures and/or strong disorder, $k_BT/\sigma_{\rm eff}\ll1$. For
$\delta=1/2$, this is precisely the Sinai type diffusion with $a=2/\delta=4$.
However, for $\delta =1$, $\langle\delta x^2(t)\rangle\approx x_{\rm in}^2[(k_B
T/\sigma_{\rm eff})\ln(t/t_0)]^2$.

Next, with increasing temperature and time, when the unity becomes negligible in
Eq.~(\ref{central}), we find the intermediate behavior
\begin{eqnarray}
\label{intermed}
\langle \delta x^2(t)\rangle\approx x_{\rm in}^2 [t/t_0]^{\alpha(t)}
\end{eqnarray}
with the time dependent scaling exponent
\begin{eqnarray}\label{intermed2}
\alpha(t)=2(k_B T/\sigma_{\rm eff})^{1/\delta}\ln^{1/\delta-1}(t/t_0).
\end{eqnarray}
This result predicts a power-law subdiffusion with a gradually changing anomalous
diffusion exponent. It should be mentioned here that this result follows from the
approximation $\exp[f(t)]-1\approx\exp[f(t)]$, for a logarithmically growing $f(t)$,
which is not fully accurate in our numerics. Nevertheless, it predicts the correct
law for $\alpha(t)$, namely, $\alpha(t)\propto \log t$, for $\delta=1/2$, see below.
This result is especially insightful for $\delta=1$ (or $\log x$ scaling): here
$\alpha(t)= 2k_BT/\sigma_{\rm eff}=\alpha_{\rm min}$, and we observe subdiffusion
of power law form with a constant anomalous diffusion exponent. Moreover, given 
the results depicted in Fig.~\ref{Fig1}, where initially $\delta>1$, intermediately
$\delta=1$, and asymptotically $\delta=1/2$, one can predict that initially
$\alpha(t)$ will diminish and reach a minimum, and then logarithmically increase
in time. This is what we actually see in the simulations with $\sigma_{\rm eff}
\approx(1.24\div1.52)\sigma$ within the temperature range $\sigma/4<k_BT<\sigma/2$,
see below. Such a behavior may indeed mistakingly be attributed to continuous
time random walk subdiffusion.

\subsection{Transient lack of ergodicity}

Due to the decaying correlations of the Gaussian potential fluctuations in our
model, with increasing temperature for a given root mean squared amplitude
$\sigma$ (decreasing $\beta\sigma$) and for times longer than a certain
crossover time, a transition to normal diffusion will gradually emerge. 
The corresponding characteristic anomalous diffusion length $L_{\rm erg}$ can be
found from the condition that the variance of the random process $w(x)=\exp(-\beta
U(x))$ equals the mean, which yields the following implicit equation for this, so
far unknown, characteristic length $L_{\rm erg}$ \cite{GoychukPRL14},
\begin{eqnarray}
\label{cond}
\int_0^1\left (1-y\right )e^{\beta^2\sigma^2g(L y)}dy=1.
\end{eqnarray}
The transition to normal diffusion starts at $\langle\delta x(t)^2\rangle\gtrapprox
L_{\rm erg}^2$. However, this transition may last extremely long, and anomalous
diffusion features may persist for appreciably long times for $|\delta x|>L_{\rm
erg}$.

For the case of linearly decaying correlations, Eq.~(\ref{cond}) can be solved
exactly, and we obtain ($L_{\rm erg}>\lambda$)
\begin{eqnarray}
\nonumber
L_{\rm erg}&=&\frac{\lambda}{(\beta\sigma)^2}\Big\{(e^{\beta^2\sigma^2}-1)-(\beta
\sigma)^2+\Big(2(\beta\sigma)^4+4(\beta\sigma)^2\\
&-&2(\beta\sigma)^2e^{(\beta\sigma)^2}+3-4e^{(\beta\sigma)^2}+e^{2(\beta\sigma)^2
}\Big)^{1/2}\Big\}
\label{L_linear}
\end{eqnarray}
which for $\beta\sigma\gtrapprox2.5$ is approximated by the simple expression
\begin{eqnarray}
\label{L_approx}
L_{\rm erg}\approx\frac{2\lambda}{(\beta\sigma)^2}e^{(\beta\sigma)^2}
\end{eqnarray}
with a very good accuracy. For other models of $g(x)$, Eq.~(\ref{cond}) is solved
numerically, the results being listed in Table \ref{Table} for several values of $\beta\sigma$.
One can see that the shortest length $L_{\rm erg}$ occurs for the linearly decaying
correlations and the longest one for the power law correlations. Somewhat
surprisingly, for the Gaussian decay the non-ergodicity length is larger than for
exponentially decaying correlations.

\begin{table}
\begin{ruledtabular}
\begin{tabular}{|p{1.3cm}|p{1cm}|p{1cm}| p{1cm}| p{1cm}|p{1.2cm}|}
$g(x)$ & $\beta\sigma=2$ & $\beta\sigma=3$  & $\beta\sigma=4$  & $\beta\sigma=5$
& $\beta\sigma=10$   \\
\hline
$e^{-|x|}$ & $34.85$ & $2070$ & $1.2\times10^{6}$ &  &  \\ 
\hline
$e^{-x^2}$ & $51.89$ & $5027$ &  &  &  \\
\hline
$\frac{1}{(1+x^2)^{\gamma/2}}$ $\gamma=0.8$ & $166.83$ & $9670$  & $6.86 \times
10^{6}$ & $4.28 \times 10^{10}$ & $7.64 \times 10^{42}$ \\
\hline
$1-|x|$ & $24.59$ & $1798$ & $1.11 \times 10^{6}$ & $5.76 \times 10^{9}$  & 
$5.37 \times 10^{41}$  \\
\hline
$e^{\beta^2\sigma^2}$ & $54.59$ & $8103$ & $8.89 \times 10^{6}$ &  $7.20
\times 10^{10}$ & $2.69 \times 10^{43}$ \\
\end{tabular}
\end{ruledtabular}
\caption{Non-ergodicity length $L_{\rm erg}$ from Eq.~(\ref{cond}), in units of
$\lambda$ and values for $e^{\beta^2\sigma^2}$.
\label{Table}}
\end{table}

As mentioned above, there exist several crossover times in the dynamics. The
first one corresponds to the transition from Sinai like diffusion to the power
law diffusion regime. The corresponding transition time $\tau^*$ can be roughly 
estimated from the condition that the argument of the exponential function in
Eq.~(\ref{central}) reaches unity. From this, $\tau^*\sim t_0\exp[\sigma_{\rm
eff}/k_BT]$. It grows exponentially fast with $\sigma_{\rm eff}/k_BT$. The second
transition time $\tau_{\rm erg}$ can be estimated from the condition of how long
the power law diffusion regime with anomalous diffusion exponent $\alpha_{\rm
min}\approx 2k_BT/\sigma_{\rm eff}$ will last.
Thus we find the
conditions $\langle x^2(t)\rangle\sim x_{\rm in}^2(\tau_{\rm erg}/t_0)^{\alpha_{
\rm min}}\sim L_{\rm erg}^2$, 
from which with $\alpha_{\rm min}\approx2k_BT/\sigma_{\rm eff}$ and Eq.~(\ref{L_approx}),
we obtain the estimate
\begin{eqnarray}
\label{tau_erg}
\tau_{\rm erg}\sim t_0\left(\frac{4\lambda^2}{x_{\rm in}^2}\frac{(k_BT)^4}{\sigma^4}
\right)^ {\frac{\sigma_{\rm eff}}{2k_BT}}\exp\left[\frac{\sigma^2\sigma_{\rm eff}}{
(k_BT)^3}\right].
\end{eqnarray}
 Notice the super-exponential growth of $\tau_{\rm erg}$ with $\sigma/
(k_BT)$. This is precisely the reason why the power law subdiffusion regime can
last so long already for moderate values $\sigma/(k_BT)\sim4\div5$, and no
transition to the regime of normal diffusion was revealed in the concrete cases
studied in Refs.~\cite{Romero98,KhouryPRL,SimonPRE13}.

\section{Results of numerical simulations}

We now proceed by checking our theoretical predictions against results of
extensive numerical simulations, finding remarkable agreement. To this end,
let us consider a continuous-space Brownian dynamics governed by the
overdamped Langevin equation  
\begin{eqnarray}
\label{langevin}
\eta\frac{dx(t)}{dt}=-\frac{dU(x)}{dx}+\sqrt{2k_BT\eta}\times\zeta(t), 
\end{eqnarray} 
where $\eta$ is the frictional coefficient and $\zeta(t)$ is unbiased, white
Gaussian noise with $\delta$-correlation $\langle\zeta(t)\zeta(t')\rangle=
\delta(t-t')$. Distance is scaled in units of $\lambda$, time in units of
$\tau_0=\lambda^2\eta/\sigma$, and temperature in $\sigma/k_B$. Initially,
$10^4$  particles are uniformly distributed in random potentials (10
realizations), and the particle motion is integrated using periodic boundary
conditions with a very large period $L_{\rm max}$. The random potential is
generated on a lattice with spacing $\Delta x=0.02$, and discrete points are
connected by parabolic splines, that is, the potential is locally parabolic,
and the corresponding force entering the Langevin equation (\ref{langevin}) is
piece-wise linear. In this respect our setup is similar to the one considered
in Refs.~\cite{Romero98,SimonPRE13,GoychukPRL14}, but different from that in
Refs.~\cite{HanesJPCM,HanesPRE}, where a discrete hopping dynamics in both
space and time was studied. In most simulations, we employ $L_{\rm max}=2^{19}\Delta
x\approx1.0485 \times10^4\lambda$. We use the stochastic Heun method with a
time integration step $\Delta t=2\times10^{-4}$ in most simulations.

\subsection{Exponential correlations}

\begin{figure*}
\includegraphics[width=8cm]{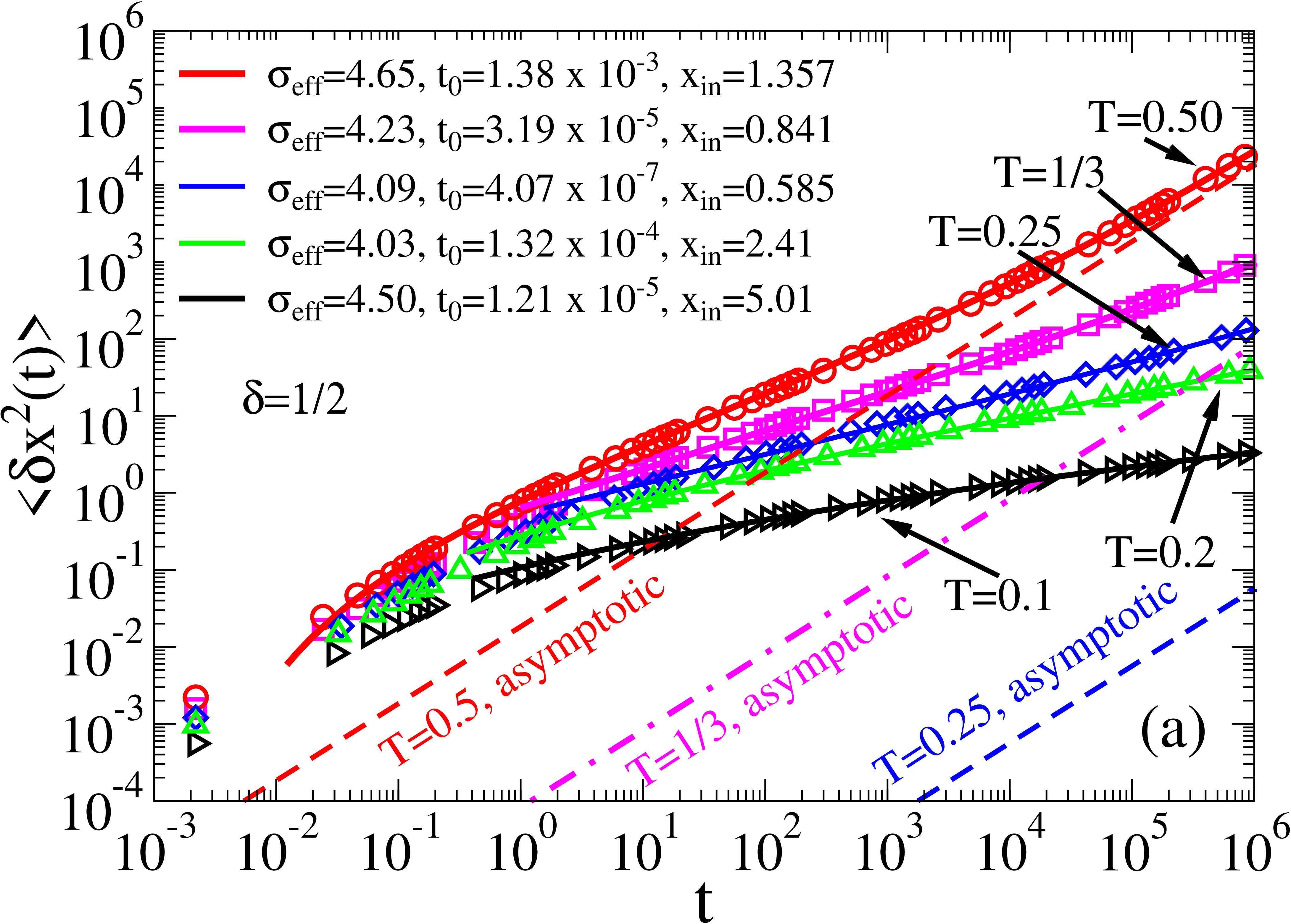}
\hspace*{0.8cm}
\includegraphics[width=8cm]{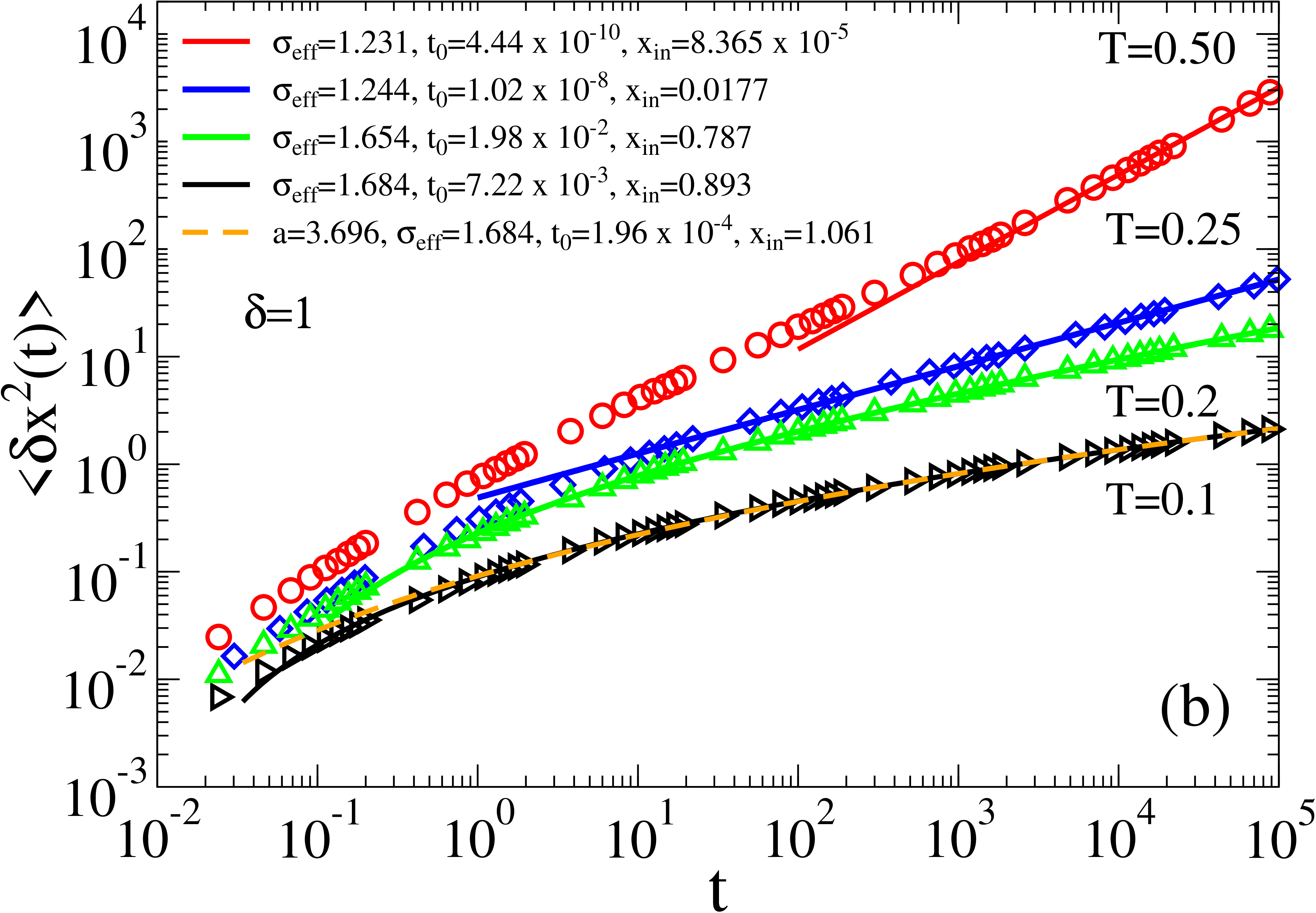}\\[0.8cm]
\includegraphics[width=8cm]{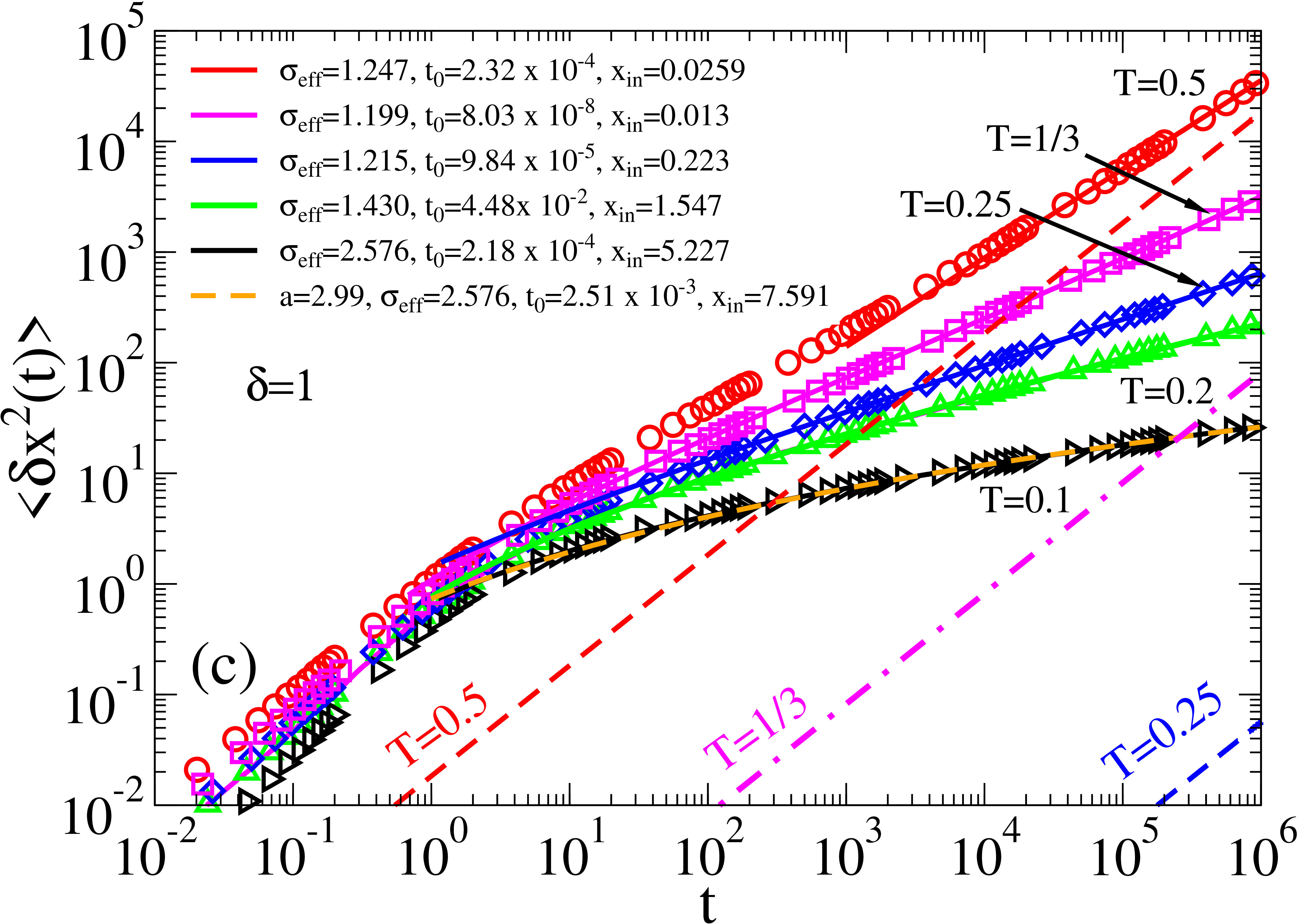}
\hspace*{0.8cm}
\includegraphics[width=8cm]{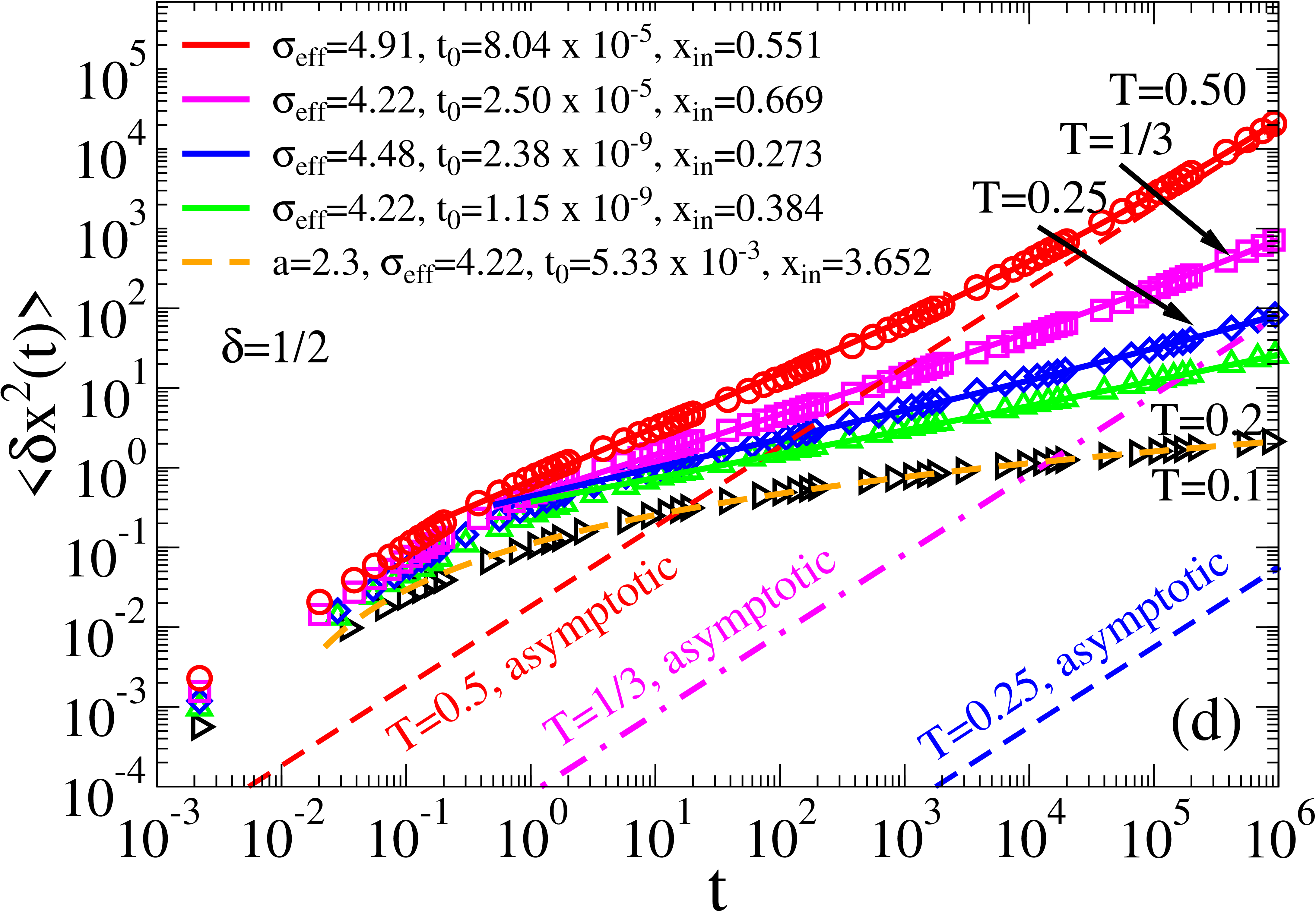}\\[0.8cm]
\includegraphics[width=8cm]{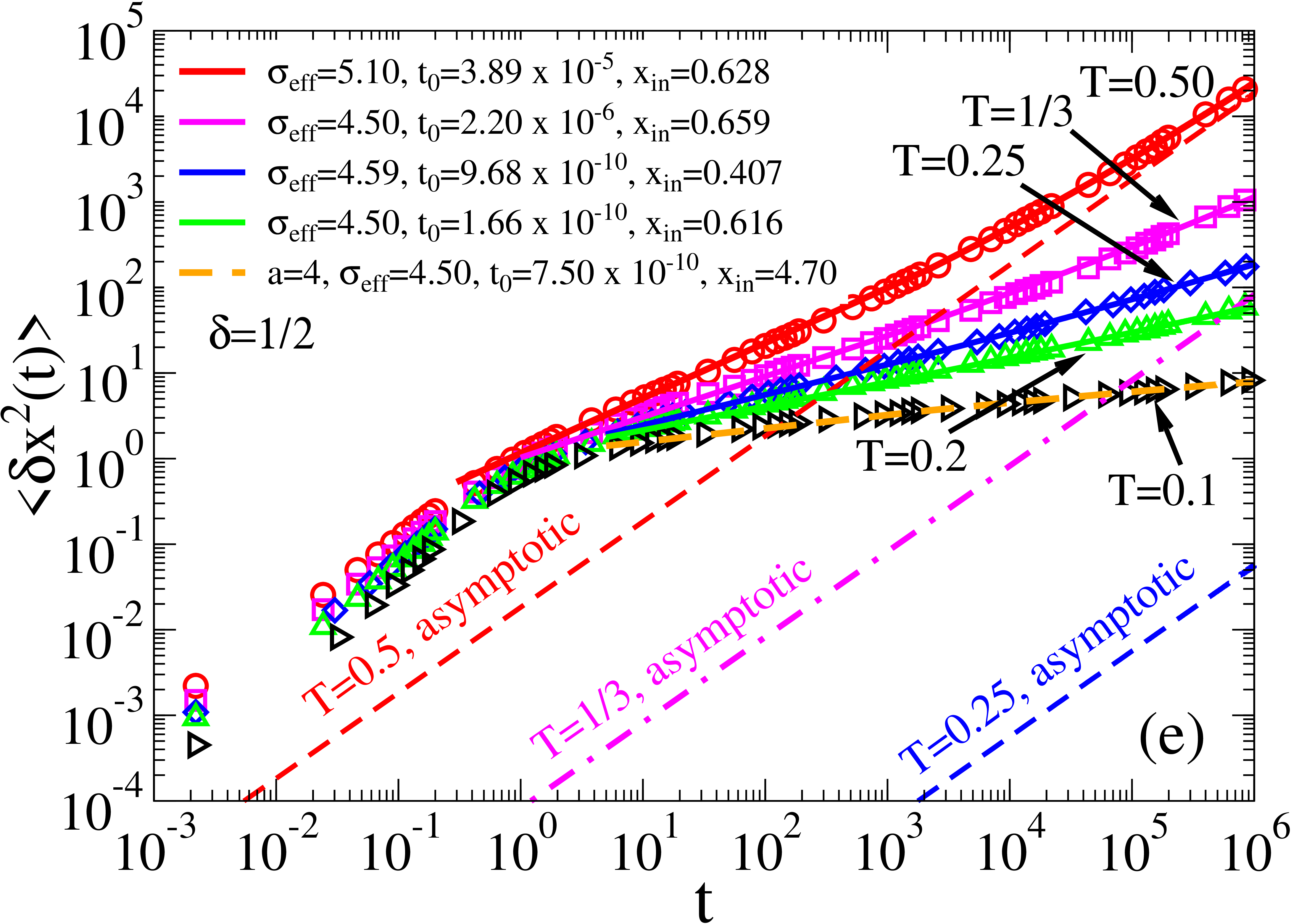}
\caption{(Color online) Ensemble-averaged mean squared displacement (symbols) for
different values of $k_BT$ in units of the disorder strength $\sigma$, for the
four different models of spatial correlation decay: (a) and (b) exponential
decay; (c) power-law decay with $\gamma=0.8$; (d) linear decay; (e) Gaussian
decay. The fit of the numerical results (full lines) is performed expression
(\ref{central}) with $\delta=1/2$ in (a), (d), and (e); and $\delta=1$ in (b) and
(c). The dashed orange line in (b) to (d) shows an alternative fit with the
generalized Sinai diffusion expression (\ref{genSinai}) and $a\approx3.70$ in (b),
$a\approx 2.99$ in (c), $a\approx 2.3$ in (d), and $a\approx 4 $ in (e). For $T=
0.1$, Sinai diffusion covers typically at least six decades of time. Even for $T=
0.5$ the fit with Eq.~(\ref{central}) and $\delta=1/2$ turns out to be very good
over 6 decades of time, as seen in panels (a), (d), and (c). For $\delta=1/2$,
$\sigma_{\rm eff}$ is typically in the range, $\sigma_{\rm eff}\sim4\div5.1$,
whereas for $\delta=1$ it typically ranges in $\sigma_{\rm eff}\sim1.2\div1.7$.
Distances are measured in units of $\lambda$ and time in units of $\tau_0=
\lambda^2\eta/\sigma$.}
\label{Fig2}
\end{figure*}

\begin{figure*}
\includegraphics[width=8cm]{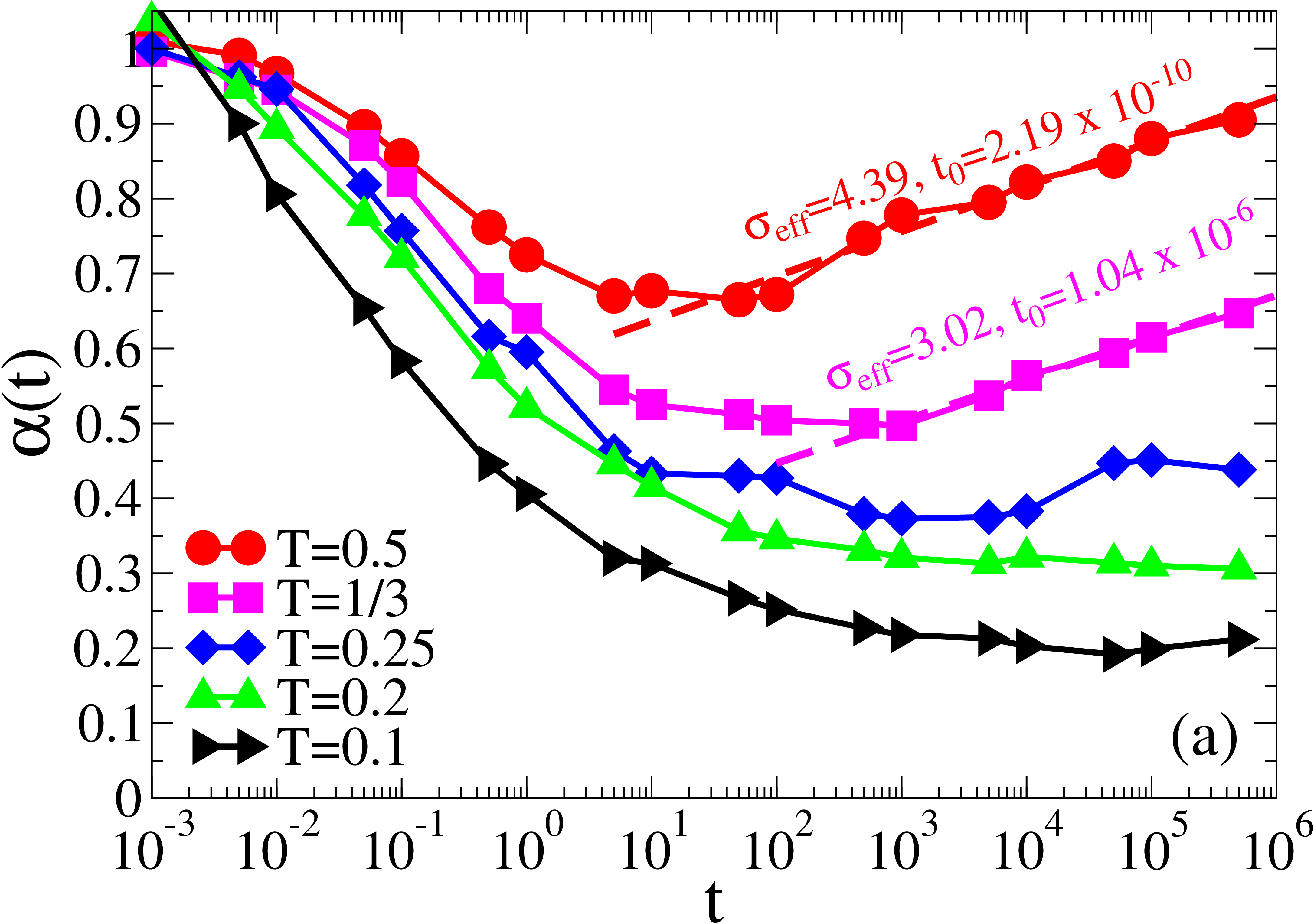}
\hspace*{0.8cm}
\includegraphics[width=8cm]{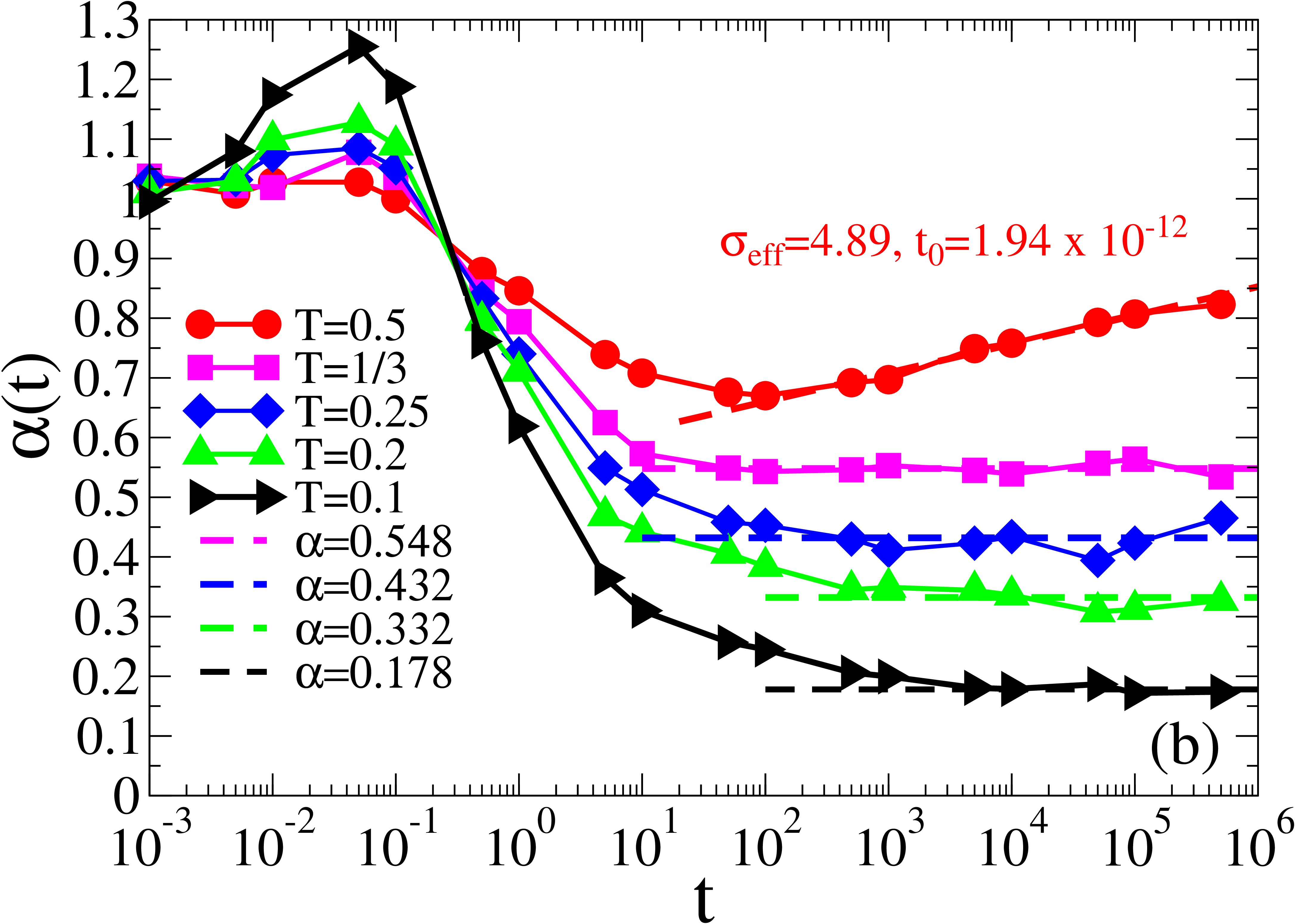}\\[1.2cm]
\vfill 
\includegraphics[width=8cm]{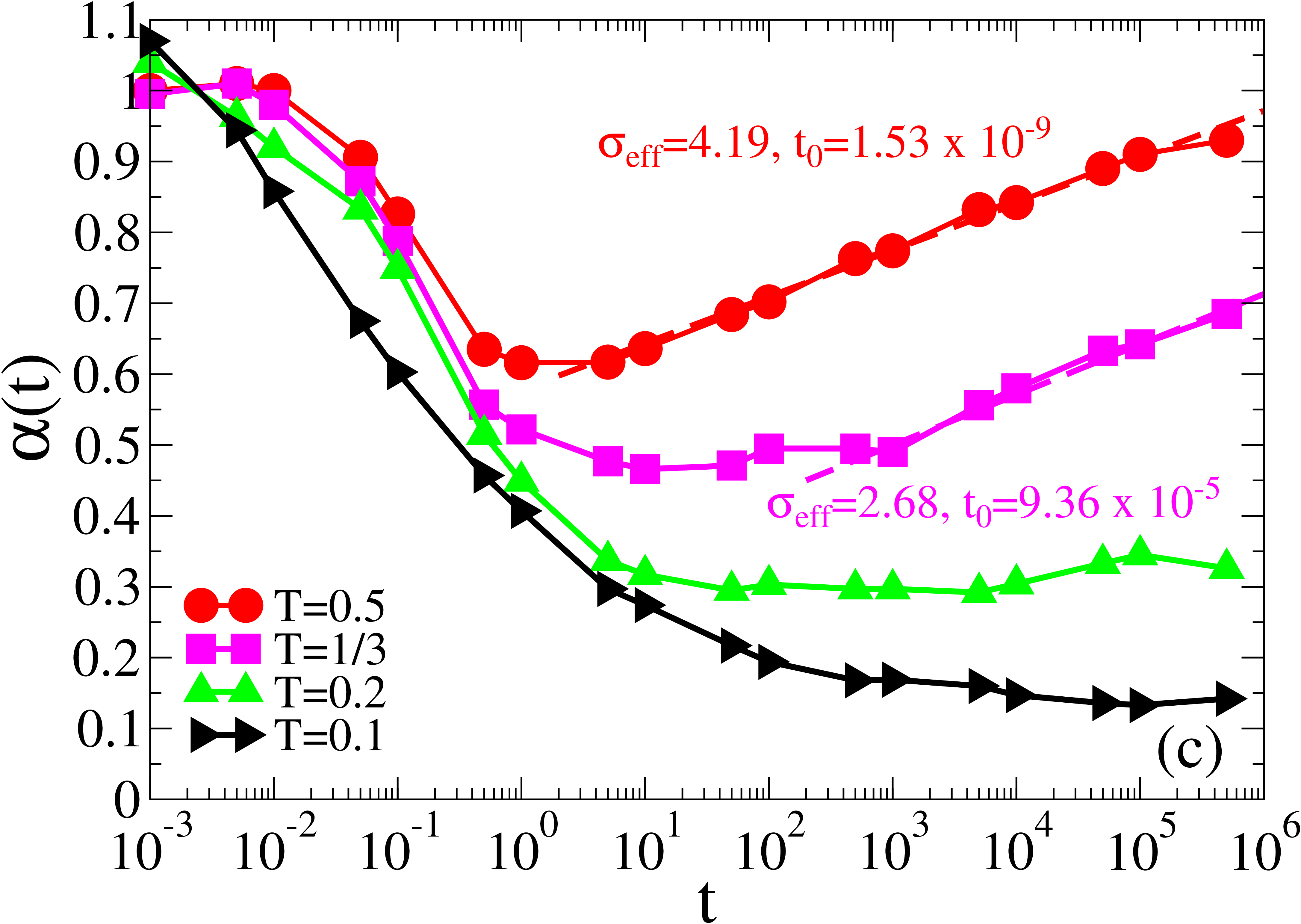}
\hspace*{0.8cm}
\includegraphics[width=8cm]{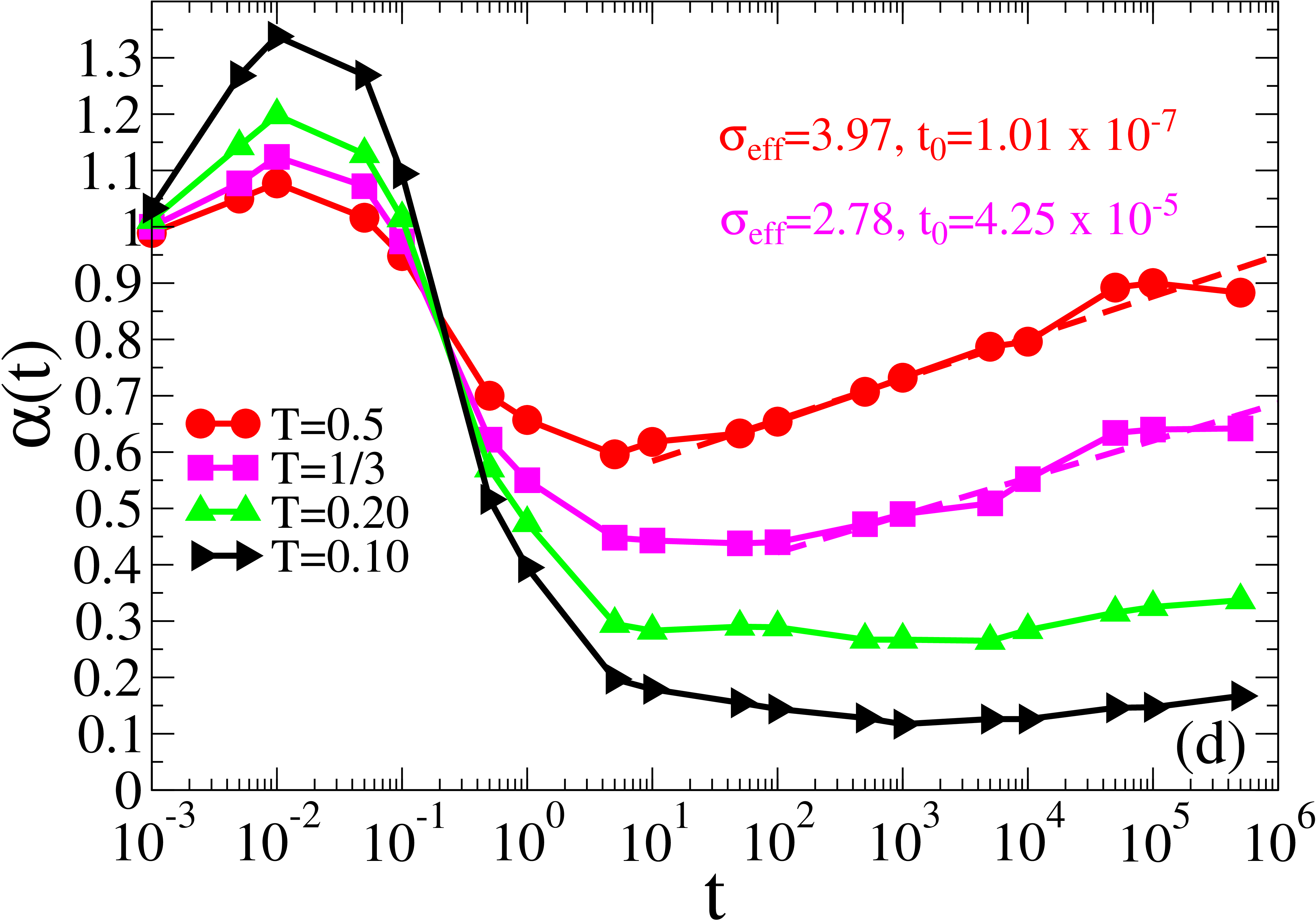}
\caption{(Color online) Time-dependent power law exponent $\alpha_{\rm eff}(t)$
for an assumed subdiffusive law $\langle\delta x^2(t)\rangle\simeq t^{\alpha(t)}$
obtained as the logarithmic derivative of the traces in Fig.~\ref{Fig2}, for
different temperatures in the case of (a) exponential correlations, (b) power law
correlations with $\gamma=0.8$, (c) linearly decaying correlations, and (d)
Gaussian correlation decay. Symbols connected by lines correspond to numerical
results, while the dashed lines are fits to the dependence $\alpha(t)=2(k_BT/
\sigma_{\rm eff})^2\ln(t/t_0)$. Fitting parameters shown in the plot.}
\label{Fig3}
\end{figure*}

\begin{figure}
\vspace*{0.4cm}
\includegraphics[width=8cm]{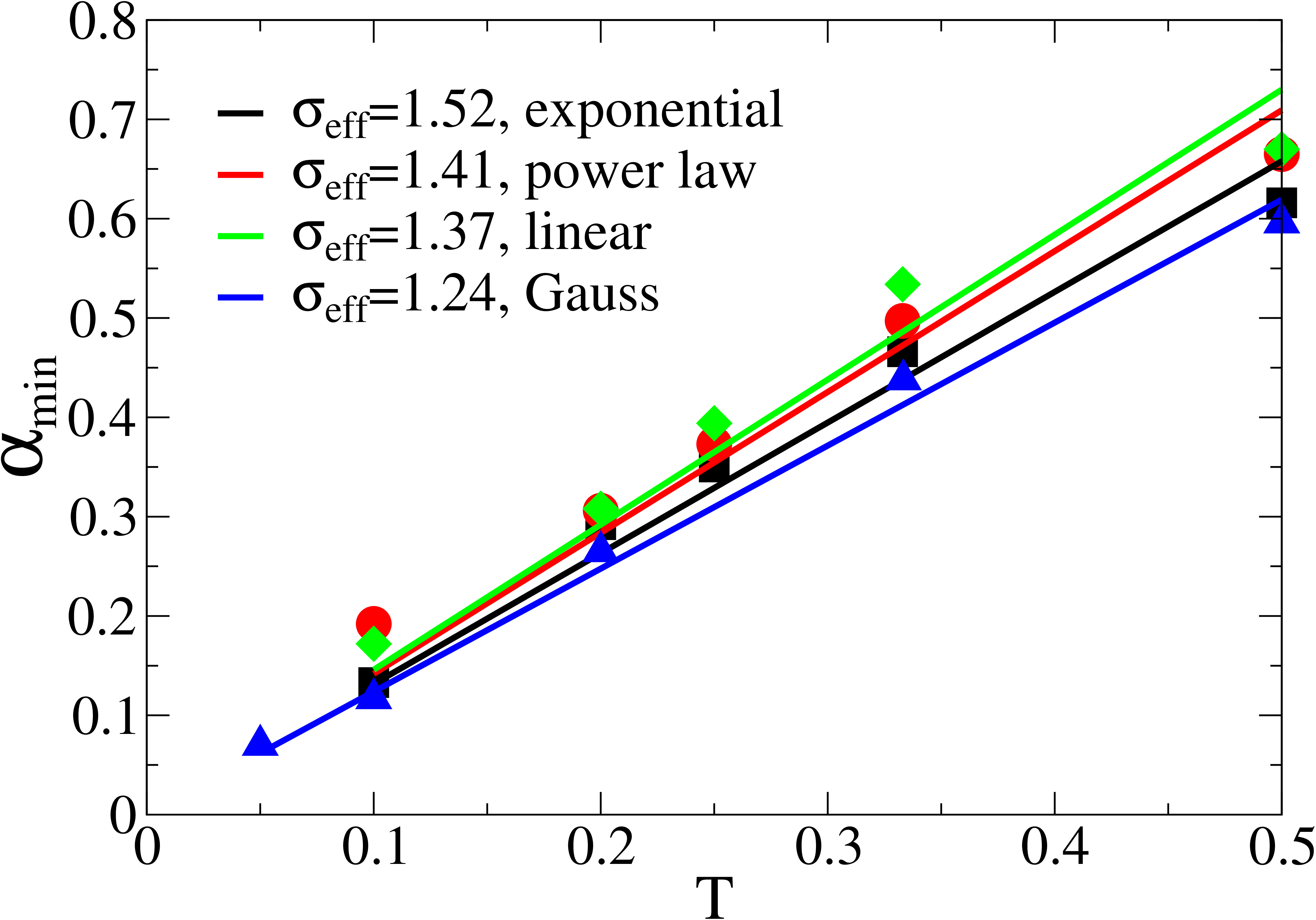}
\caption{(Color online) Dependence of $\alpha_{\rm min}$ on temperature $T$ for
the four different models of spatial correlations discussed in this paper, and
the corresponding best linear fits with the dependence $\alpha\approx2k_BT/
\sigma_{\rm eff}$. The values of the corresponding $\sigma_{\rm eff}$ are shown
in the plot.}
\label{Fig4}
\end{figure}

\begin{figure*}
\includegraphics[width=8cm]{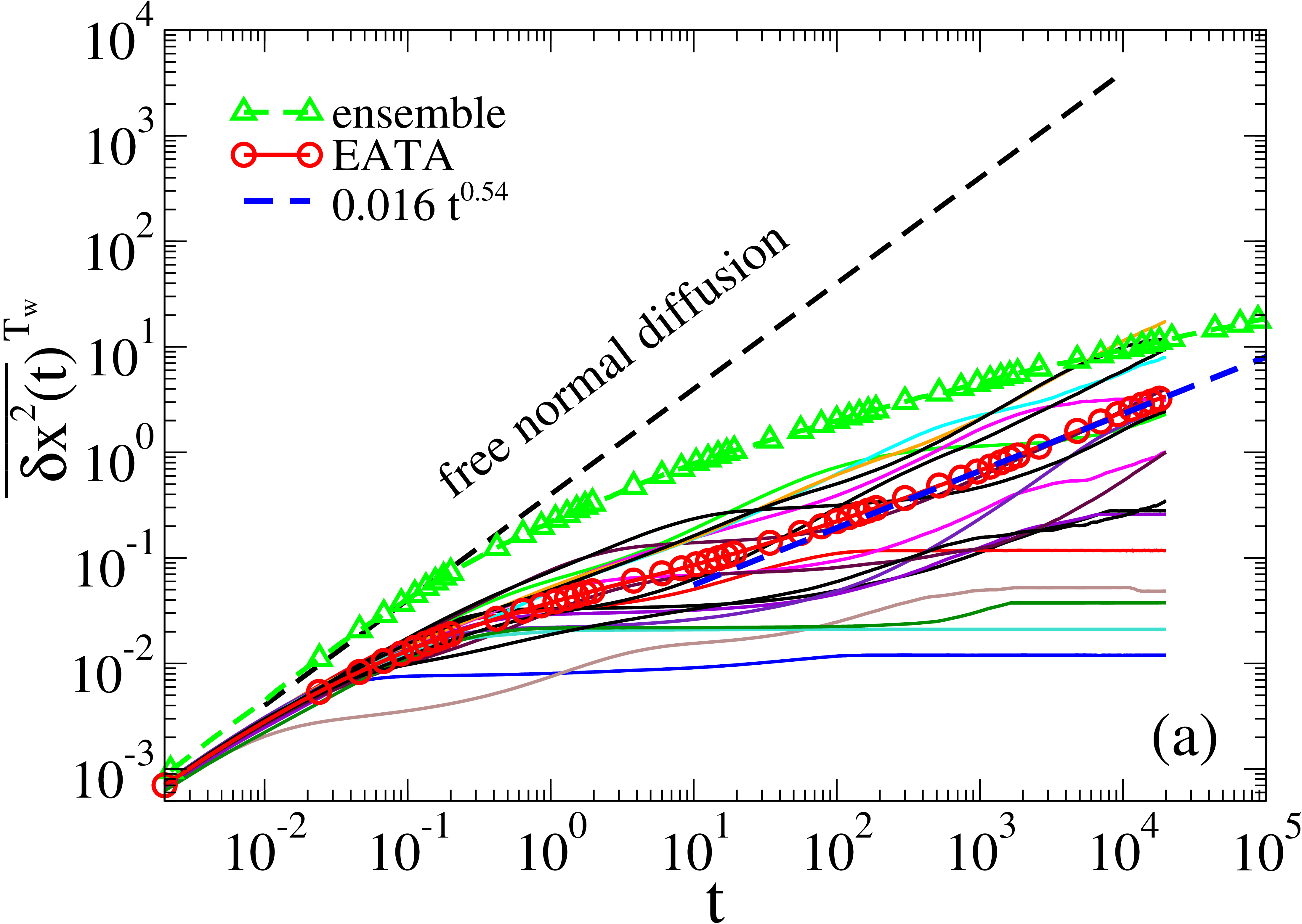}
\hspace*{0.8cm}
\includegraphics[width=8cm]{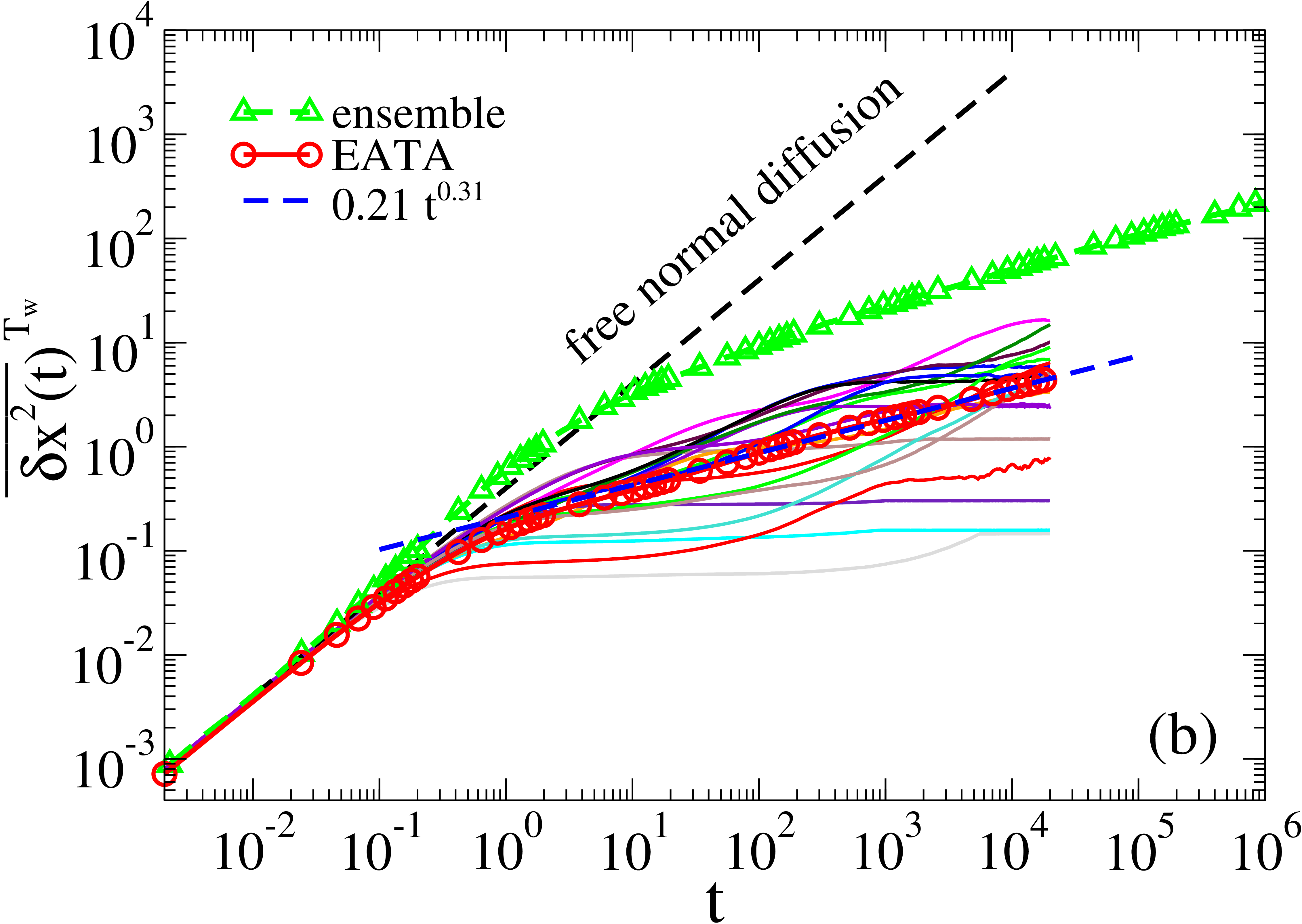}\\[0.8cm]
\includegraphics[width=8cm]{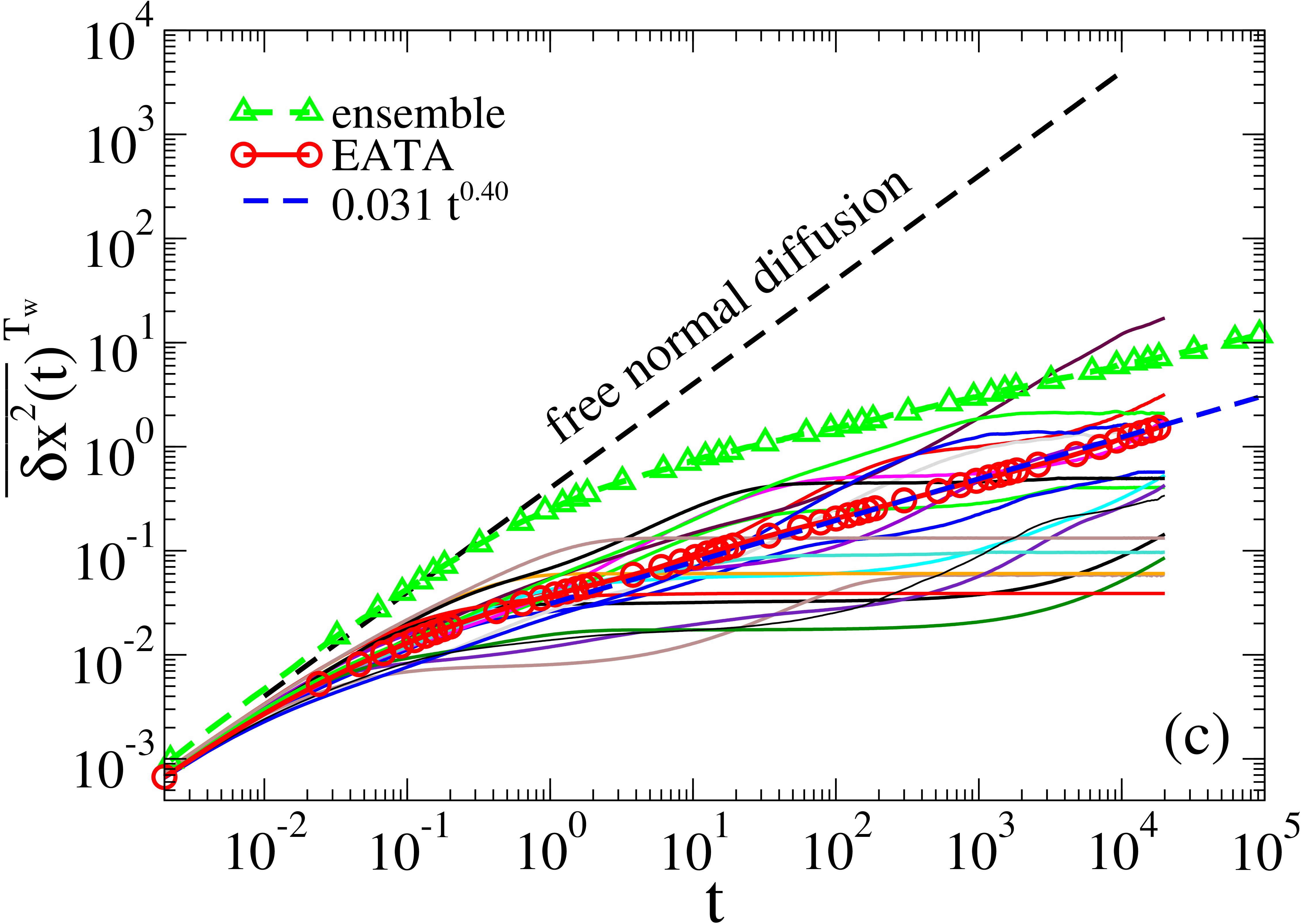}
\hspace*{0.8cm}
\includegraphics[width=8cm]{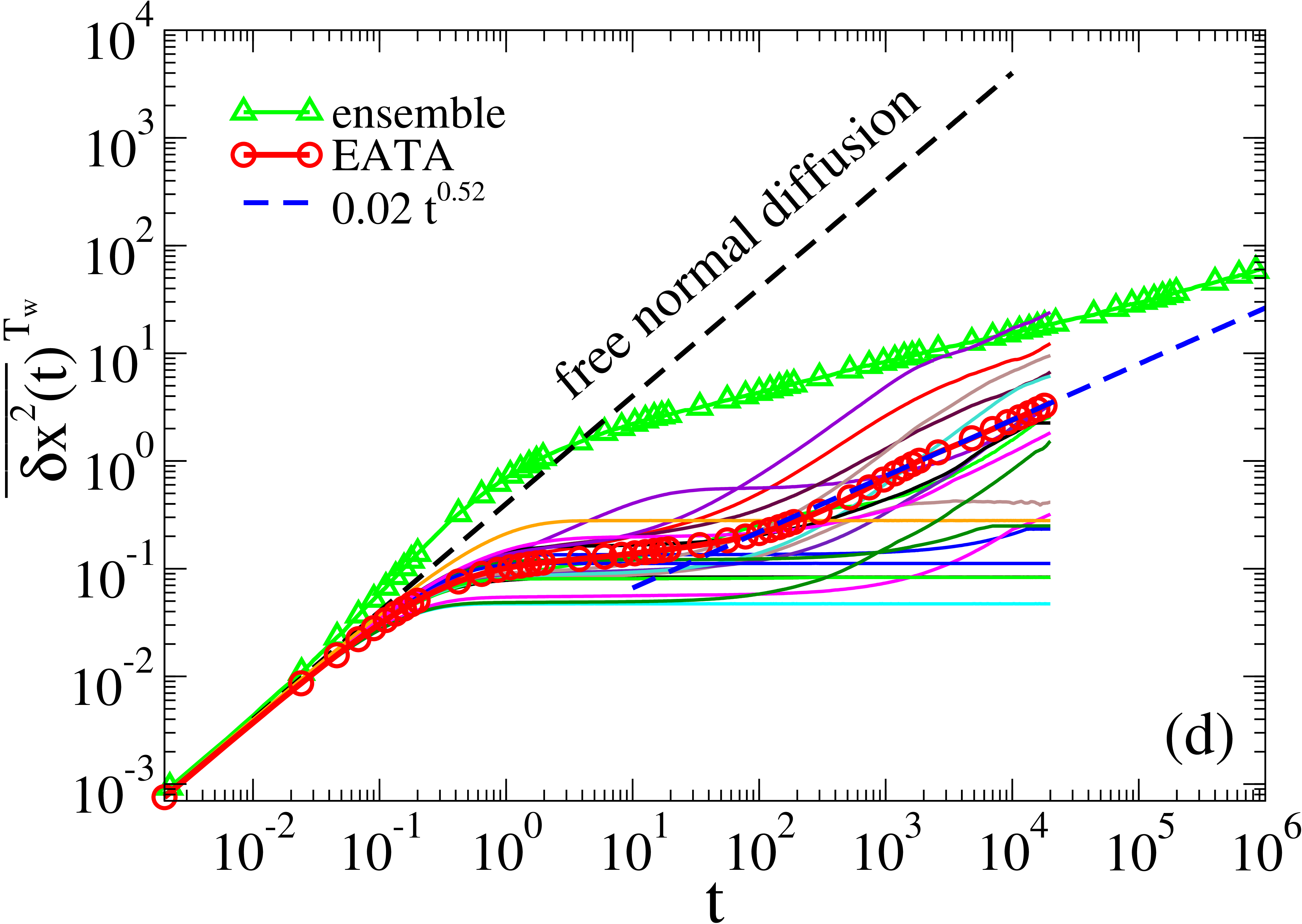}
\caption{(Color online) Single-trajectory, time averaged mean squared displacement
for $T=0.2$ and (a) exponential correlations, (b) power-law correlations with
$\gamma=0.8$, (c) linearly decaying correlations, and (d) Gaussian correlation
decay. The time averaging window (measurement time) was $T_{\rm w}=2\times10^6$.
20 trajectory averages are depicted in each case starting from different locations.
The results of the ensemble-averaged, as well as ensemble-averaged time-averaged (EATA), and
free normal mean squared displacements are also depicted for comparison. The EATA
mean squared displacement is nicely fitted by a power-law dependence, which is also
shown in the plot. The asymptotic limit of the disorder-renormalized  normal diffusion
cannot be plotted for comparison because  $D_{\rm ren}=\exp(-25)D_0\approx1.39\times10^{-11}D_0$,
with $D_0=0.2$.  Subdiffusion is by many orders of magnitude
faster.} 
\label{Fig5}
\end{figure*}

We start our numerical analysis with the case of the exponentially decaying
correlations, $g(x)=\exp(-|x|)$. From the simulated trajectories, we first
perform an ensemble averaging to obtain the mean squared displacement. The
results are depicted in Fig.~\ref{Fig2} in panels (a) and (b). In Fig.~\ref{Fig2}a,
the fit of the numerical results is based on Eq.~(\ref{central}) using $\delta=1/2$,
$\sigma_{\rm eff}\sim4\div4.7$, and $x_{\rm in}\sim0.59\div5.01$. Note that the
fitting values $\sigma_{\rm eff}$ differ only by a factor of less than two from
the theoretical value $2.83$ in Fig.~\ref{Fig1}. The corresponding theoretical
value $x_0$ is $x_0\approx 0.31$. Here, the agreement is worse, however, the
numerical results in Fig.~\ref{Fig1} in this case are also somewhat different
from the theoretical asymptotics, which is still not reached. The same numerical
data are fitted to expression (\ref{central}) with $\delta=1$ in Fig.~\ref{Fig2}b,
with $\sigma_{\rm eff}\sim1.23\div1.69$. For $T=0.1$ and $T=0.2$, both fits have
almost the same quality, although the fit with $\delta=1$ appears slightly better.
For $T=0.1$, Sinai like diffusion (\ref{genSinai}) with $a\approx3.7$, near to
the Sinai value $a=4$, covers about six decades in time, Fig.~\ref{Fig2}b.
It is worthwhile noting that for $T\leq 0.25$ the fit with $\delta=1$ works better,
see Fig.~\ref{Fig2}b, due to the initial $\log x$ scaling in Fig.~\ref{Fig1} (see
symbols therein). For this reason, $\alpha(t)$ shows a nearly constant behavior,
at intermediate temperatures, for an extended time period, Fig.~\ref{Fig3}a. In
fact, in this regime the power law exponent is nearly proportional to temperature, 
$\alpha\approx2k_BT/\sigma_{\rm eff}$, where the value of $\sigma_{\rm eff}$ is
obtained from fitting the minimal value $\alpha_{\rm min}$ in Fig.~\ref{Fig3}a by
this dependence. It turns out that $\sigma_{\rm eff}\approx1.52$, Fig.~\ref{Fig4}.
However, with increasing temperature, when diffusion covers larger distances, the
fit with $\delta=1/2$ becomes much better, compare the case $T=0.5$ in
Figs.~\ref{Fig2}a and \ref{Fig2}b. An excellent fit holds over about six to seven
decades in time with $\delta=1/2$, where the explicit time dependence of the
anomalous diffusion exponent $\alpha(t)$ in Eq.~(\ref{intermed}) becomes apparent
for sufficiently long times. At larger temperatures, for instance, $T=1$, the
crossover to the asymptotic regime of normal diffusion is already accomplished
up to $t_{\rm max}=10^6$ in our simulations. Such high temperatures, or weak
disorder strengths, are not of interest for anomalous diffusion of the kind
discussed. Generally $\alpha_{\rm eff}$ reaches a minimum $\alpha_{\rm min}$,
may stay nearly constant for a certain time period, and then logarithmically
increases, a dependence which can be fitted by Eq.~(\ref{intermed2}) with
$\delta=1/2$ and some values of $\sigma_{\rm eff}$ and $t_0$, which are different
from those used in Fig.~\ref{Fig2}a, see Fig.~\ref{Fig3}a for $T=0.5$ and $T=1/3$.
The reason for this discrepancy in the corresponding fitting values is that a
transition from Eq.~(\ref{central}) to Eqs.~(\ref{intermed}) and (\ref{intermed2})
is still not quite justified numerically. Nevertheless, the prediction of
a logarithmically increasing $\alpha(t)$, as confirmed by the simulations, is a
remarkable success of our simple scaling theory.

Interestingly, our simulations demonstrate that generally the observed subdiffusion
is much faster than the corresponding limit of disorder-renormalized normal
diffusion, which is also shown in Fig.~\ref{Fig2}a for several values of the
temperature. The presence of correlations thus leads to a dramatic increase in
the particle mobility on intermediate but relevant time scales. Note that whereas
for $T=0.5$ this limit is already gradually approached in Fig.~\ref{Fig1}, the
corresponding asymptotics cannot even be depicted for $T=0.2$ and $T=0.1$ in this
figure, as they lie outside of the plotting range, and, therefore, are completely
irrelevant on the corresponding time scale.

As discussed above, the physical origin of the observed subdiffusion is due to
a weak breaking of ergodicity. Hence, single trajectory time averages
\begin{equation}
\overline{\delta x^2(t)}^{\rm T_w}=\frac{1}{{\rm T_w}-t}\int_0^{{\rm T_w}-t}\Big[
\delta x(t|t')\Big]^2dt'
\end{equation}
of the mean-squared displacement $\delta x(t|t')=x(t+t')-x(t')$ over the time
window ${\rm T_w}$  are expected to be very different from the above ensemble
result, and their amplitudes should be broadly scattered \cite{HePRL08,MetzlerPCCP}.
This is indeed so, as evidence in Fig.~\ref{Fig5}a, where at all times $t\ll{\rm
T_w}$, and hence this scatter is not a trivial statistical effect occurring when
$t\sim{\rm T_w}$. Remarkably, some of the particle trajectories are quickly
localized, while others are diffusing very fast. The slow ones start near to and
become trapped in a low potential valley, while the fast particles start from a
relatively high value of $U(x)$ and move downhill to a much lower value of $U(x)$,
thus experiencing a local energy bias. We note that this phenomenon is analogous to
the Golosov localization in standard Sinai diffusion \cite{Golosov84,Bouchaud1990},
when particles starting at the same (thermal) initial position are not significantly
separated in the course of time. One may observe two such very close trajectories
in Fig.~\ref{Fig5}a. Thus, particles diffuse similarly and are correlated. This
feature can be very important for the diffusion of proteins on DNA, which may be
locally biased, even if the bias is absent on average. This behavior is very 
different from the scatter of single trajectory averages in the case of annealed
continuous time random walk subdiffusion with divergent mean resident time. In the
latter case, even identical particles starting at the same place will follow very
different, diverging trajectories. This feature can therefore be used for a crucial
experimental test to distinguish between different types of subdiffusion. 
Overall, both ensemble and
time averaged mean squared displacements are by many orders of magnitude faster
than in the limit of renormalized normal diffusion whose diffusion coefficient
is suppressed by the factor $\exp(-25)\approx1.39\times 10^{-11}$, as compared
with free normal diffusion in Fig.~\ref{Fig5}a for the same temperature $T=0.2$.

As can be seen from single trajectory recordings (not shown), particles typically
continue their diffusive motion after being localized for a certain time. This
feature can also been seen in some trajectory averages depicted in Fig.~\ref{Fig5}a,
where the diffusional spread displays a step-like feature. Namely, the diffusional
spread continues after temporally reaching a plateau. In data analyses this
might mistakingly be attributed to a continuous time random walk subdiffusion
behavior. The ensemble average of the single trajectory time averages is also of
interest. This is how experimentalists often proceed to smooth out single trajectory
averages \cite{MetzlerPCCP}. Such an ensemble averaged time average (EATA) is
also shown in Fig.~\ref{Fig5}. It is nicely fitted by a power-law time dependence.
Hence, on the level of this ensemble averaged time averaged mean squared
displacement the power law subdiffusive regime is established earlier.

\begin{figure*}
\includegraphics[width=8cm]{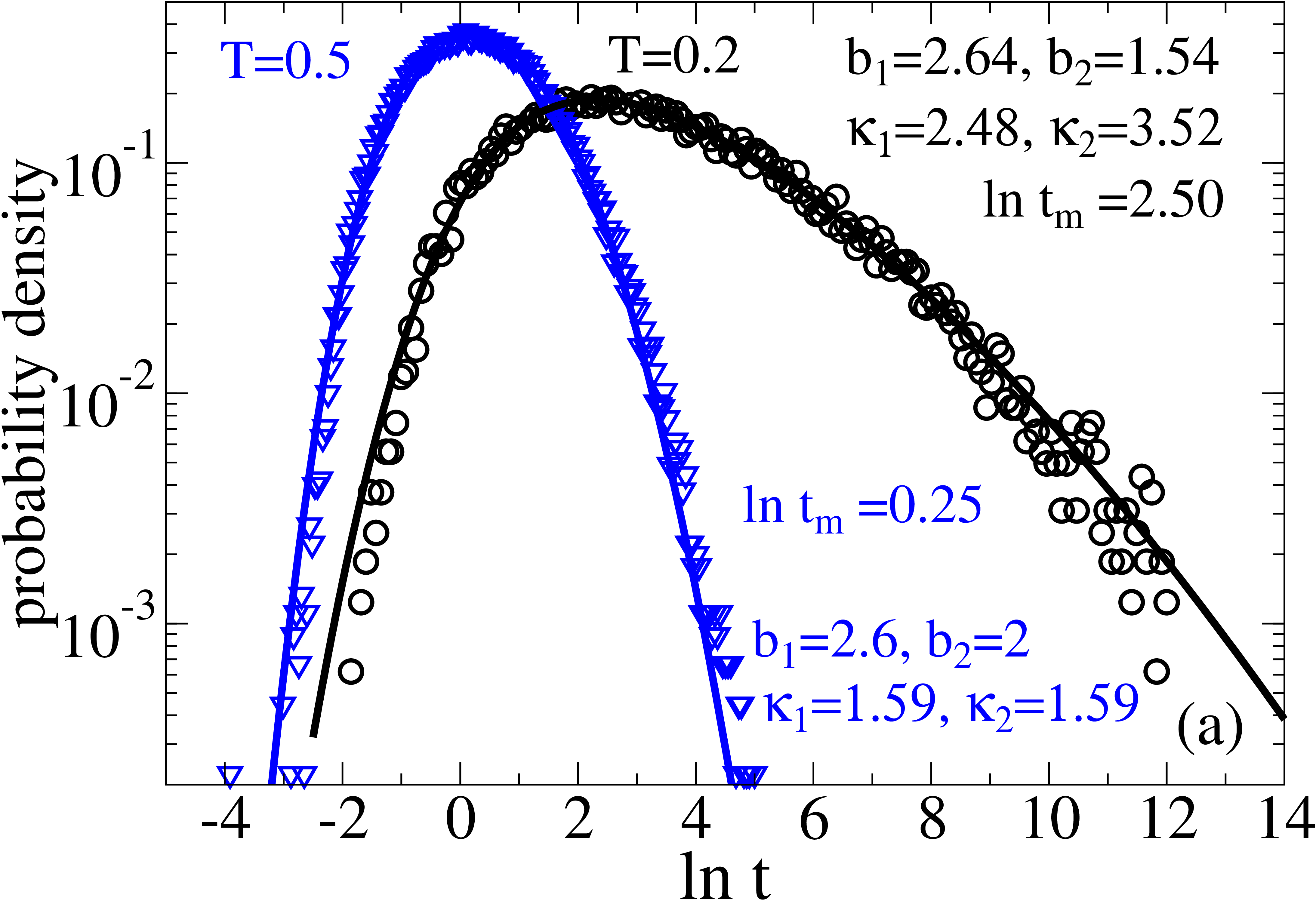}
\hspace*{0.8cm}
\includegraphics[width=8cm]{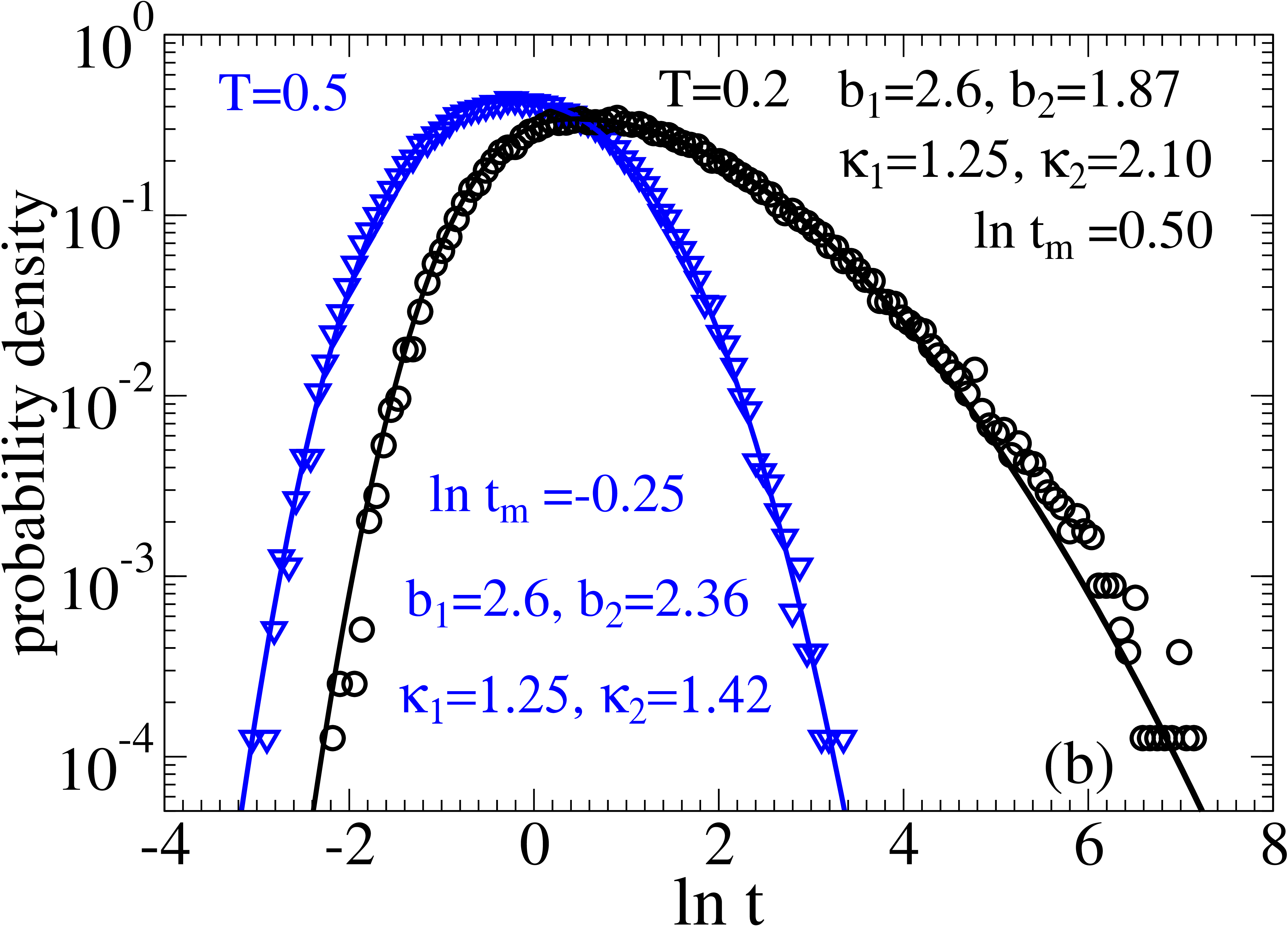}\\[0.8cm]
\includegraphics[width=8cm]{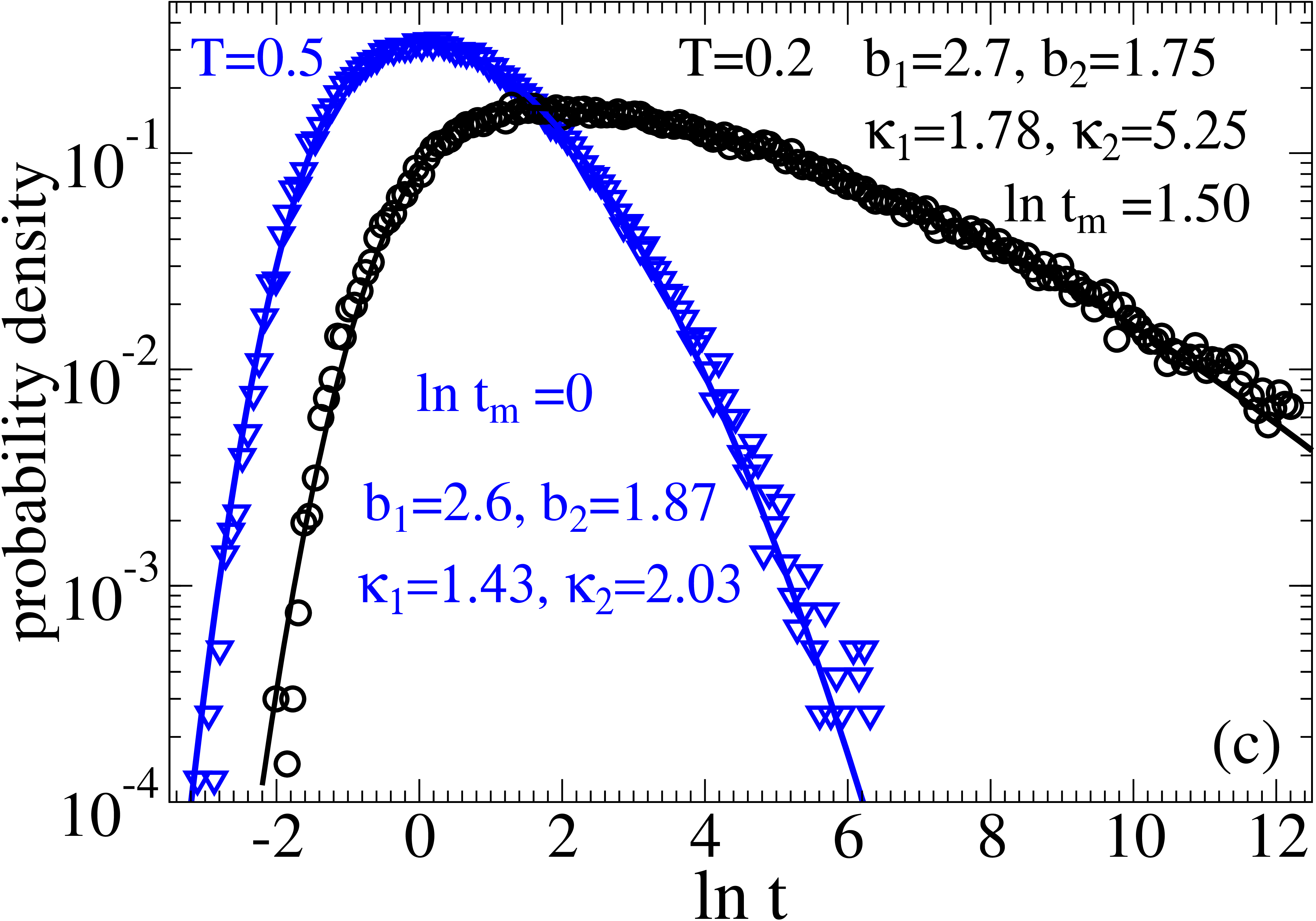}
\hspace*{0.8cm}
\includegraphics[width=8cm]{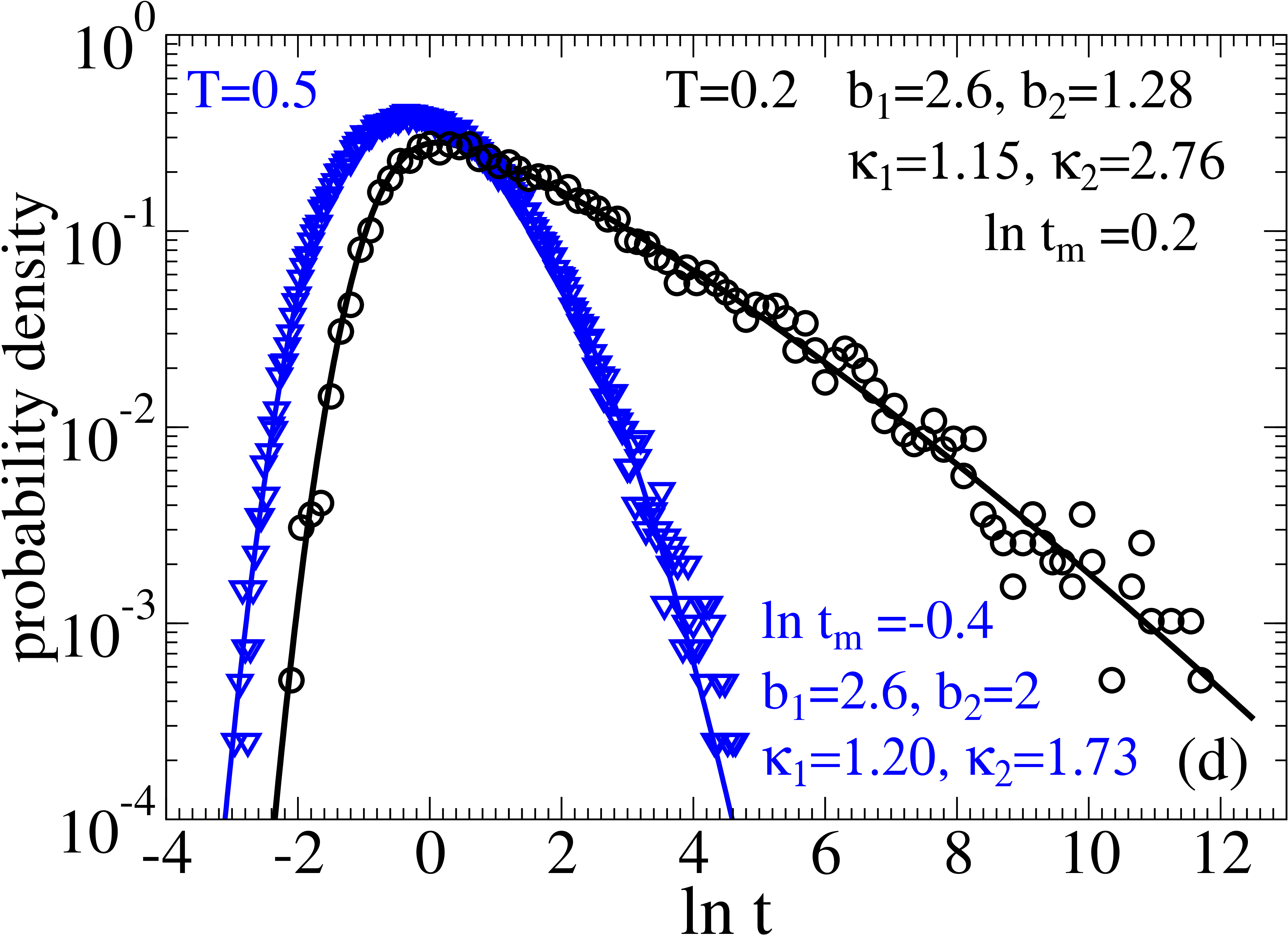}
\caption{(Color online) Probability density function of log-transformed first
escape times from the interval $[-\lambda,\lambda]$ for two temperatures $T=0.5$
(in blue) and $T=0.2$ (in black). The cases in the different panels are (a)
exponential correlations, (b) power-law correlations with $\gamma=0.8$, (c)
linearly decaying correlations, and (d) Gaussian correlation decay. The symbols
represent simulations data, the lines correspond to a fit with the probability
density (\ref{distro}). Parameters are shown in the panels. With decreasing
temperature the distributions become broader and the parameter $b_2$ smaller.}
\label{Fig6}
\end{figure*}

All moments of the corresponding residence time distribution in any finite spatial
domain are finite. Moreover, the average residence time is much (for $k_BT\ll\sigma$
orders of magnitude) smaller than the one expected from the normal diffusion
characterized the renormalized result $D_{\rm ren}$ \cite{GoychukPRL14}. In the present case, weak
ergodicity breaking does not rely on infinite mean residence times, what makes it
especially interesting in biological applications. Residence time distributions of
escape times for particles initially located in the center of a symmetric interval
$\pm \lambda$ in space for $k_BT\leq0.5\sigma$ are nicely described by a
generalized log-normal distribution reading
\begin{eqnarray}
\nonumber
\psi(t)&=&\frac{C}{t}\Big[ e^{-|\ln(t/t_m)/\kappa_1|^{b_1}}\theta(t_m-t)\\
&&+e^{-|\ln(t/t_m)/\kappa_2|^{b_2}}\theta(t-t_m)\Big],
\label{distro}
\end{eqnarray}
where $C=b_1b_2/[b_2\kappa_1\Gamma(1/b_1)+b_1\kappa_2\Gamma(1/b_2)]$ is a
renormalization constant, $b_{1,2}>1$, and $\kappa_{1,2}>0$. $\Gamma(x)$
denotes the complete Gamma function. For any $b_{1,2}>1$, this distribution
(\ref{distro}) has finite moments $\langle t^n\rangle$ of all orders $n$. 
For $b_1=b_2=2$ and $\kappa_1=\kappa_2$, this is the well-known log-normal
distribution. However, in our case $b_1$ and $b_2$ are different, and $\kappa_1
\neq\kappa_2$. Namely, for the exponential correlations $b_1\approx2.6\div2.7$,
and it is weakly temperature dependent. However, $b_2$ does depend visibly on
temperature. For $T=0.5$, $b_2=2$, and it becomes smaller with decreasing
temperature while $t_m$ becomes longer. For instance, for $T=0.2$, $b_2\approx
1.54$, see Fig.~\ref{Fig6}a, where the distribution of times is plotted for
the log-transformed variable $\ln t$, attaining a maximum at $\ln t_m$.

\subsection{Power law correlations}

We proceed with the case of power law correlations of the spatial potential
fluctuations, with $\gamma=0.8$. The ensemble averages for the mean squared
displacement are depicted in Fig.~\ref{Fig2}c. Here, the fit with $\delta=1$
works generally better, which can be rationalized from Fig.~\ref{Fig1}, and
for $T=0.1$ a generalized Sinai diffusion with $a\approx3$ nicely fits the
numerical data over six decades in time. Note in Fig.~\ref{Fig3}b that
$\alpha(t)$ stays nearly constant over five decades in time, up to the end
of the simulations even for $T=1/3$. For the larger $T=0.5$, it also grows
logarithmically in time, as in the case of exponential correlations.  Why
$\alpha(t)$ changes slower in time in this case and a fit with $\delta=1$
and $\sigma_{\rm eff}\approx1.41$ works nicely over many time decades for
intermediate temperatures can be rationalized from the fact that the
non-ergodicity length $L_{\rm erg}$ in this case is much larger, see Table
\ref{Table}. The case of power law correlations is especially important for
the diffusion of regulatory proteins on DNA strands, where such long-range
correlations with $\gamma\sim0.6\div0.8$ emerge due to the way biological
information is encoded in the base pair sequence. Our results imply that
such diffusion should be typically anomalously slow with $\alpha\sim 1.41\;k_BT/
\sigma$ in the corresponding range of temperatures, for $\gamma=0.8$. It must
be stressed in this respect that diffusion does not become immediately normal
for $\langle \delta x^2(t)\rangle>L_{\rm erg}^2$, but rather a very slow
crossover with growing $\alpha(t)$ emerges, see Fig.~\ref{Fig2}c for $T=0.5$.
Here, the transition to the asymptotic regime of normal diffusion is still far
from being established.

Somewhat surprisingly, for power law correlations
subdiffusion is essentially faster in absolute terms than in the case of
exponentially short correlations, see Fig.~\ref{Fig7}. This may first appear
counter-intuitive. However, one reason for this feature becomes clear from the
potential fluctuations depicted in Fig.~\ref{Fig1}, which for the exponential
correlations are essentially larger (thinking in units of $k_BT$). Clearly,
exponential correlations and other singular (in the limit $\Delta x\to 0$)
models of static disorder present a preferred case to observe Sinai type
diffusion in such systems. The second explanation is that in the case of
power law correlations a local bias is present on a much longer scale. This
leads to an essential acceleration of diffusion in the ensemble sense, see
Fig.~\ref{Fig5}b, where the ensemble mean squared displacement is in fact
superdiffusive (see also Fig.~\ref{Fig3}b for the initial regime) within several
correlation lengths at $T=0.2$. Remarkably, it is even faster than the potential
free diffusion. This is namely due to the presence a local bias, which is also
the reason for the Golosov phenomenon in Sinai diffusion. The single trajectory
time averaged mean squared displacement, however, does not show such a striking
feature, see Fig.~\ref{Fig5}b. The reasoning for this feature is that the local
bias is averaged out in the single trajectory time averaging for a sufficiently
large measurement time $T_{\rm w}$. Single trajectory averages spread out slower
than the ensemble averaged result and, nevertheless, their broadening is many
orders of magnitude faster than the result of the renormalized normal diffusion.
This is why even for $\sigma\sim 4\div5$, relevant for some important biophysical
situations \cite{GerlandPNAS,LassigReview}, regulatory proteins can diffuse on
DNA tracks despite of the fact that the classical de Gennes-Zwanzig-B\"assler
result would predict that they should be practically localized (on biophysically
relevant time scales) for such a strong disorder. Naturally, single trajectory
averages exhibit a large scatter due to a transient lack of ergodicity. It is
important to mention that a strong single trajectory scatter was indeed observed
experimentally \cite{WangPRL}. Note also that such a local bias can be functionally
very important, directing the protein toward a specific binding site on DNA.
The ensemble averaged time averaged mean squared displacement in this case also
shows a power law scaling in time, see Fig.~\ref{Fig5}b. Judging from the power
law exponent $\alpha_{\rm EATA}=0.31$, it appears to be slower than in the case
of exponential correlations, where $\alpha_{\rm EATA}=0.54$, compare with
Fig.~\ref{Fig5}a. However, the prefactor in this case is, in fact, much larger,
which makes it in fact faster rather than slower.

The distribution of escape times is also nicely captured by the generalized
log-normal distribution (\ref{distro}), as seen in Fig.~\ref{Fig6}b. Note
that the escape times in this case are essentially shorter than in the case
of exponential correlations, Fig.~\ref{Fig6}a. Particularly, the scaling
exponent $b_2$ is larger and the half width $\kappa_2$ is smaller. Conversely,
the power exponent $b_1$ is pretty robust, $b_1\sim2.6$ (also for other models
of disorder, see below).

\subsection{Linear correlations}

The case of linear correlations presents another important model of singular
disorder with potential fluctuations growing with diminishing $\Delta x$.
Therefore, it is expected to be similar to the case of exponential correlations,
despite the fundamental difference between these two models. The non-ergodicity
length $L_{\rm erg}$ here is the shortest within the four considered models, see
Table \ref{Table}. The ensemble averaged mean squared displacement is depicted
in Fig.~\ref{Fig2}d. Qualitatively, it appears very similar to the previous cases,
however, it turns out to be the slowest one, as Fig.~\ref{Fig7} reveals. In this
case, $\sigma_{\rm eff}\sim4.22\div4.91$, and $x_{\rm in}\sim 0.27\div0.67$,
except from $T=0.1$, where $x_{\rm in}\approx3.65$. Now, the agreement with $x_0
\approx0.314$ is much better. It might seem paradoxical that the shortest
correlations, which exactly vanish at distances exceeding $\lambda$, yield the
slowest diffusion. However, it must be kept in mind that in the absence of
spatial correlation we have just normal diffusion, which is orders of magnitude
slower than the considered correlation-induced subdiffusion. The time behavior
of the anomalous diffusion exponent $\alpha(t)$ is indeed more similar to the
one in the case of exponential correlations than to the case of power law
correlations, see Fig.~\ref{Fig3}c, and compare with panels (a) and (b) therein.
The scatter of the single trajectory time averaged mean squared displacements
is also quite pronounced, see Fig.~\ref{Fig5}c. The escape times obey the same
distribution (\ref{distro}), as depicted in Fig.~\ref{Fig6}c, but with different
parameters. Similar to previous cases, $b_1\sim2.6\div2.7$, nearly independent
of temperature, while $b_2$ strongly depends on temperature, decreasing with
decreasing temperature. All temporal moments of the resident times remain,
however, finite with decreasing temperature.

\subsection{Gaussian correlations}

\begin{figure}
\includegraphics[width=8cm]{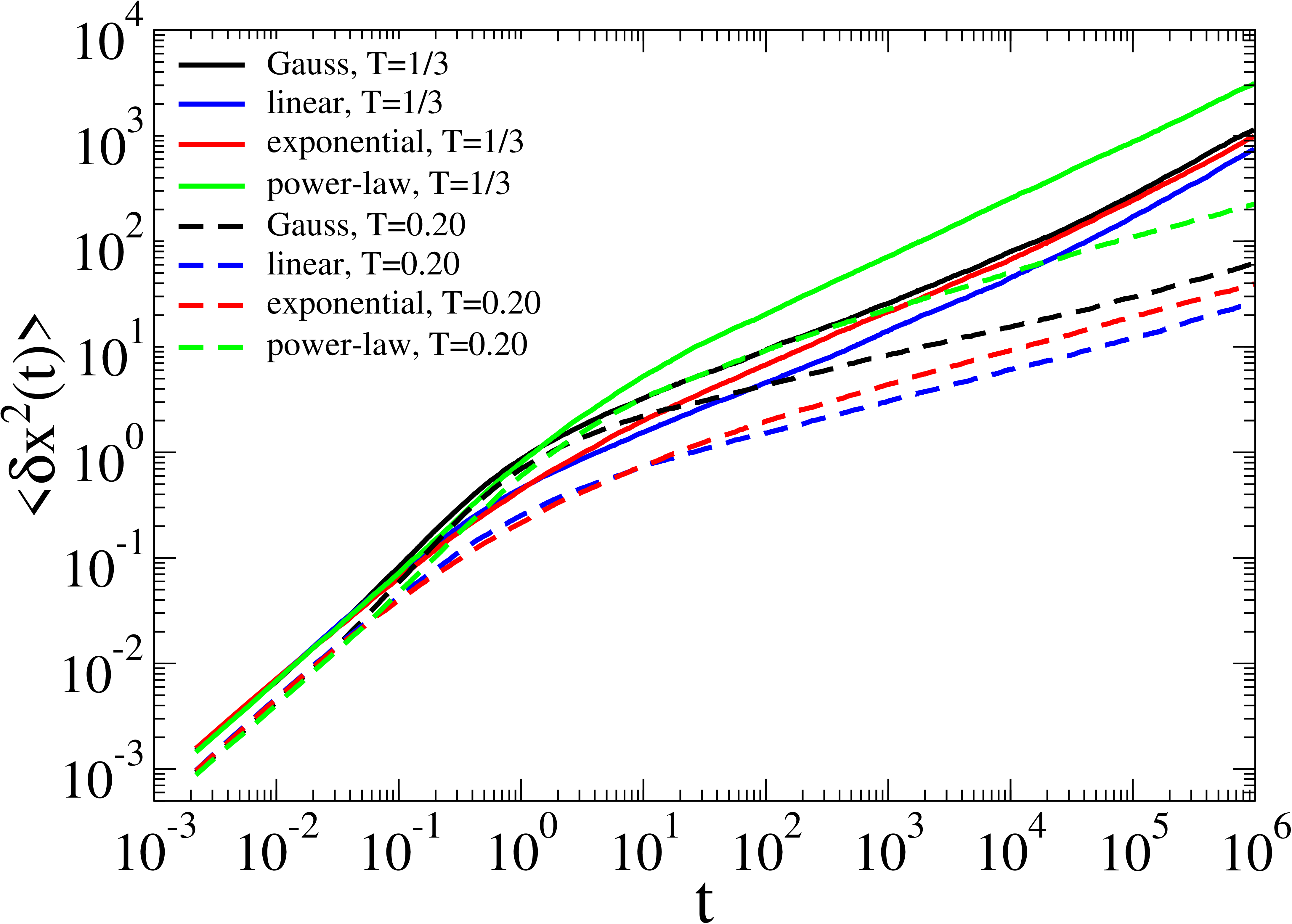}
\caption{(Color online) Comparison of the ensemble-averaged mean squared
displacement for the four different models of the correlation decay and for
two values of temperature. Seemingly against intuition, the mean squared
displacement is the largest in the case of power-law spatial correlations,
where the spatial non-ergodicity scale is the longest. Moreover, diffusion
is slowest in the case of linear correlations, where the non-ergodicity length
is the shortest in all four cases, see Table \ref{Table}.}
\label{Fig7}
\end{figure}

Finally, we consider the case of Gaussian spatial correlations. Judging from
Fig.~\ref{Fig1} it should be the second fastest of the four models considered
here. This is indeed the case, as seen in Fig.~\ref{Fig7}. The ensemble averaged
mean squared displacements display very similar generic features as in the other
cases, see Fig.~\ref{Fig2}e. Here, $\sigma_{\rm eff}\sim4.50\div5.10$ and $x_{\rm
in}\sim0.41\div4.70$, whereas the theoretical is $x_0\approx 2.22$. Once again,
the discrepancy by a factor of less than two with the theoretical $\sigma_{\rm
eff}=2.83$, and at least the correct order of magnitude for $x_{\rm in}$ are
quite impressive, given the simplicity of our scaling argumentation. Sinai type
diffusion with $a=4$ is featured in the low temperature behavior at $T=0.1$.
In this case, however, $\alpha(t)$ already starts to slightly increase at the
end of the simulations already at $T=0.1$, Fig.~\ref{Fig3}d. This type of
correlations indeed presents the worst case to observe a Sinai like diffusion.
The reasons are quite obvious: Namely, (1) this model of disorder is not
singular and (2) the correlations are short-ranged. Single trajectory averages
of the mean squared displacement are strongly scattered, and their ensemble average
yields a power law dependence for sufficiently large times, Fig.~\ref{Fig5}d.
Finally, the distribution of the escape times obeys the same common pattern,
as seen in Fig.~\ref{Fig6}d.

\section{Discussion and conclusions}

Diffusion in systems characterized by Gaussian fluctuations of the potential
with spatially decaying correlations is typically thought of as normally
diffusive on experimentally relevant time scales, following the renormalization
idea for these systems developed by de Gennes, Zwanzig, and B{\"a}ssler.
Following recent discoveries of extended anomalous diffusion in such systems,
we here provided a clear physical picture for the origin of subdiffusion in
stationary Gaussian random potentials with decaying spatial correlations. While
the anomalous diffusion is, of course, transient, the scaling theory developed
and confirmed herein demonstrates that this transient may easily reach macroscopic
time scales for a physically relevant strength of the potential
fluctuations.

The primary reason for the origin of this subdiffusion is that the maximal amplitude of
the potential fluctuations grows logarithmically with the distance, in accordance
with the asymptotic theoretical result (\ref{first}) and the simulations data
depicted in Fig.~\ref{Fig1}. This behavior is universal, and it has a predictive
power. In particular, it allows to predict for which model of the potential correlations
decay the anomalous diffusion will be faster in
absolute terms---and this prediction completely agrees with numerics. The
strongest trapping effects leading to anomalous diffusion occur in the case of
singular models of disorder, such as exponential or linear decays of the
correlations, for which the stationary energy autocorrelation function is not
differentiable at the origin. This leads to unbounded static force root mean
squared fluctuations $\langle f^2\rangle^{1/2}\to\infty$ in the strict 
$\Delta x\to 0$ limit of the coarse graining. Strictly speaking, such models are not
amenable to any Langevin simulations unless the random potential is considered
on a lattice with a spatial grid $\Delta x$, as incorporated in our simulations,
with discrete potential points connected by parabolic splines (locally piece-wise
linear forces). A maximal time step of the Langevin simulations must be chosen
accordingly, depending on $\Delta x$ and the averaged local force root mean
square $\langle f^2\rangle^{1/2}$. Singular disorder models are regularized
accordingly. However, they display the largest potential fluctuations in
Fig.~\ref{Fig1}.

Spatially increasing potential fluctuations may appear strange and contradictory
to the very fact that the random potential is stationary and possesses the well
defined root mean squared magnitude $\sigma$. This property of continuous albeit
logarithmic increase with distance is, however, the very cornerstone of the theory
of extreme events \cite{Castillo}: An extreme event will always occur, if only
one waits sufficiently long, and even if the probability density of such events
is strongly localized such as in the case of the considered Gaussian distribution.
In our case, the extreme events occur in space rather than time, but have the
same origin. The law (\ref{first}) of potential fluctuations with a $(\log x)^{
\delta}$ scaling ($\delta=1/2$) is valid asymptotically. The longer the spatial
correlations of the potential fluctuations reach, the later this asymptotic is
achieved. Numerically, we also observed different scaling exponents such as
$\delta=1$ transiently and $\delta>1$ initially.

Based on these observations we put forward a very simple scaling theory of
anomalous diffusion in the potential landscape $U(x)$, which in essence is
very similar to the scaling theory of Sinai diffusion developed in the
classical Ref.~\cite{Bouchaud1990}. It leads to our major result given in
Eq.~(\ref{central}) and explains why for sufficiently low temperatures we
indeed observe a generalized Sinai diffusion with a power law exponent, which
is generally different from $a=4$ and can be even smaller as shown in
Figs.~\ref{Fig2}b, c, and d: The diffusion is formally even slower than for the
case of Sinai diffusion in random force fields. Our analytical results were
shown to be valid over five to seven decades in time, at least, which is a
remarkable success of the relatively simple scaling approach. The nominally
ultraslow diffusion is in fact many orders of magnitude faster than the
asymptotical limit of the disorder-renormalized normal diffusion, which for
the physically relevant parameters considered here simply cannot be attained
neither physically nor numerically for such a strong disorder. Hence, it becomes
completely irrelevant in such settings. This provides a striking example of how
the formally valid mathematical result of asymptotic, disorder-renormalized
normal diffusion in the de Gennes-Zwanzig-and B{\"a}ssler theory may produce
physically inappropriate descriptions on meso- and macroscales.

This result also leads to a number of further predictions, which are confirmed
in numerical simulations and agree with previous simulations of the considered
stochastic dynamics in random potentials. Namely, this kind of anomalous
diffusion is generally characterized by a time dependent power law exponent
$\alpha(t)$, for which we obtained the theoretical result (\ref{intermed2}).
This result predicts that, after an initial decay ($\delta>1$), the anomalous
diffusion exponent $\alpha(t)$ will reach a minimum for which $\alpha_{\rm
min}=2k_BT/\sigma_{\rm eff}$ (for $\delta=1$) and then gradually grow in
logarithmic fashion, $\alpha(t)=2(k_BT/\sigma_{\rm eff})^2\ln(t/t_0)$
($\delta=1/2$). These two nontrivial predictions are remarkably well confirmed
by extensive numerical simulations. Indeed, $\alpha_{\rm min}=2k_BT/\sigma_{\rm
eff}$ with $\sigma_{\rm eff}\sim1.24\div1.52$, within a broad range of
temperatures, as demonstrated in Fig.~\ref{Fig3}. Moreover, at sufficiently large
temperatures, $\alpha(t)$ indeed grows logarithmically, see Fig.~\ref{Fig3}. It
is indeed remarkable how a simple scaling theory can be predictive to such a
degree. Of course, the scaling results cannot be fully accurate on a quantitative
level and claim full consistency over the entire range of parameters and times
studied herein. In particular, the values of $\sigma_{\rm eff}$ and $t_0$ used
to fit the numerical data, for instance, in Figs.~\ref{Fig2} and \ref{Fig3}
are in fact different. This is due to the fact that the approximation of
Eq.~(\ref{central}) by Eqs.~(\ref{intermed}) and (\ref{intermed2}) is, indeed, a bit too coarse.
Nevertheless, the coherence and validity of our scaling theoretical predictions
are compelling.

We note that in the biologically relevant and important case of power law decaying
correlations the scaling $\delta=1$ applies to a much larger range of times and
temperatures, as shown in Figs.~\ref{Fig2}c and \ref{Fig3}b. In this case, power
law subdiffusion of the form $\langle \delta x^2(t)\rangle\propto t^{\alpha}$ with
$\alpha\approx1.41\; k_BT/\sigma$ in fact presents a good approximation over a broad
range of times and temperatures. This behavior may easily be confused with the law
$\alpha\approx k_BT/\sigma$ produced by the mean field continuous time random
walk with exponential energy disorder \cite{Hughes,AvrahamHavlin,Metzler01}. In
striking contrast to the latter model, however, the residence times in a finite
spatial domain are not power-law distributed but follow the generalized log-normal
distribution (\ref{distro}) shown in Fig.~\ref{Fig6}, for all the studied models
of the correlation decay. All moments of this distribution are finite, which is a
very attractive physical feature. This subdiffusion is clearly non-ergodic, as it
is demonstrated in Fig.~\ref{Fig5}, where the maximal time is merely $1\%$ of the
time window $T_{\rm w}$ used to accumulate the single-trajectory averages. Hence,
a large scatter is a truly non-ergodic effect, even if it is a transient one and
must vanish in the mathematical limit $T_{\rm w}\to\infty$ \cite{GoychukPRL14}. It
is, however, not necessarily attainable experimentally.

The origin of this lack of ergodicity was already explained in
Ref.~\cite{GoychukPRL14}: It is due to the absence of self-averaging of the
statistical weight function $w(x)=\exp[\pm\beta U(x)]$ on the spatial scale
$L_{\rm erg}$, implicitly defined in relation (\ref{cond}), depending on the
disorder autocorrelation function $g(x)$. Solving this equation analytically
in the case of linearly decaying correlations yields the exact result
(\ref{L_linear}) and a very handy and highly accurate approximate result in
Eq.~(\ref{L_approx}). In this case of linear decay, $L_{\rm erg}$ turns out to
be the shortest one within all models of $g(x)$ considered in this paper, see
Table \ref{Table}. This result implies that $L_{\rm erg}$ in units of the
correlation scale $\lambda$ is only by a factor $(1/2)(\sigma/k_BT)^2$ smaller
than the factor $\exp[\sigma^2/(k_BT)^2]$ suppressing the asymptotically normal
diffusion coefficient in the renormalization sense. This, in turn, means that
such mesoscopic subdiffusion can readily reach macroscopic scales even for a
moderate disorder strength $\sigma\sim 4\div 5 \;k_BT$ featuring many physical
and biological systems already at room temperatures.

Different from the annealed continuous time random walk subdiffusion,
the single trajectory time averages for the mean squared displacement in
the present case of quenched potential fluctuations are characterized by
different power law exponents $\alpha\neq1$ \cite{GoychukPRL14}, and not
just a linear scaling with the lag time (erroneously suggesting normal
diffusion) but with a significantly scattered diffusion coefficient
\cite{LubelskiPRL08,HePRL08}. Importantly, especially with respect to
biological applications is that the subdiffusion considered here is in fact
orders of magnitude faster than the normal diffusion predicted by the classical
renormalization result with effective diffusivity $D_{\rm ren}$. For example,
for the diffusion of regulatory proteins on DNA tracks,  $\sigma \approx
4.3\;k_BT_{\rm room}$, which can be deduced from the results presented in
Ref.~\cite{LassigReview}. Hence, $D_{\rm ren}=9.33\times 10^{-9}D_0$, and for
a typical experimental value $D_0=3\;\mu {\rm m^2/s}$ \cite{ElfScience07}
it would become $D_{\rm ren}\approx 2.8\times 10^{-2} {\rm nm^2/s}$. This
would mean that within some hundreds of seconds the diffusional spread would
be merely several nanometers. However, this renormalization result 
underestimates massively, on the relevant time scales of this phenomenon, 
the actual protein mobility. Our results demonstrate that the actually
occurring subdiffusion is orders of magnitude faster than one suggested by
this normal diffusion result of the renormalization approach. Hence,
correlations-induced persistent subdiffusion makes diffusional search
feasible in such a situation despite a strong binding-energy disorder.

An important role in applications may also be played by the presence of a local
bias, which is especially clearly expressed in the case of power law correlations.
This is the analogous reason for the Golosov phenomenon in the case of random
potential exhibiting Brownian motion in space \cite{Bouchaud1990}. Then,
the particles which were initially localized nearby diffuse similarly, in a
correlated fashion. The distance between them does not grow dramatically in
time, being bounded. One can find similar single trajectory averages also in
our numerical results. Conversely, different particles starting in locally
different environments can move into opposite directions, and this can give
rise to an enhanced diffusivity in the ensemble sense. In fact, due to this
reason an initial regime of superdiffusion can be realized in Fig.~\ref{Fig5}b
and d on the ensemble level in the case of power-law correlations, for which
it can extend over several scaling lengths $\lambda$, see also
Fig.~\ref{Fig3}b and d for small times. In the case of single trajectory time
averages, the local bias averages out. Therefore, such a superdiffusive regime
is absent. Diffusion on the level of single trajectories is typically slower
than the one on the ensemble level, see, for instance, Fig.~\ref{Fig5}.
The difference between the ensemble average and the ensemble-averaged time average
of the mean squared displacement becomes smaller with increasing time. The latter
ensemble-time-averages display a power law dependence on time for sufficiently
large times even in the Sinai like regime on the ensemble level, which can be
an important observation with respect to possible experimental manifestations.

To conclude, we elucidated the physical mechanism leading to subdiffusion in
stationary correlated potentials with spatially decorrelating Gaussian disorder,
and we showed that a generalized Sinai diffusion typically emerges at sufficiently
low temperatures and/or strong disorder for various models of decaying
correlations. Our scaling theory also explains how a standard power law
subdiffusion emerges with increasing temperature and in the course of
time. Such subdiffusion is weakly non-ergodic, displays a local bias and
proceeds much faster than de Gennes-B\"assler-Zwanzig limit of normal diffusion,
which for sufficiently low temperatures and/or finite size of the system
simply cannot be attained physically on typical mesoscales. We believe that
our results provide a new vista on the old problem of potential disorder, and
that they will be very useful in the context of non-ergodic diffusion processes,
especially in relation to various biologically relevant problems on the cellular
level. Likely they are also important for diffusion of colloidal particles in
laser created random potentials, a conjecture calling for further experimental
studies of such systems.

\section*{Acknowledgment} 
Funding of this research by the Deutsche Forschungsgemeinschaft (German
Research Foundation), Grant GO 2052/3-1 is gratefully acknowledged.

\bibliographystyle{PRX}
\bibliography{aps}

\end{document}